\documentclass[twocolumn, tighten]{aastex62} 

\usepackage{amsmath}
\usepackage{amssymb}
\usepackage{mathtools}
\usepackage{mathrsfs}

\usepackage{morefloats}
\usepackage{bm}
\usepackage{enumerate}
\usepackage{booktabs}
\usepackage{graphicx}

\usepackage{soul}

\usepackage[makeroom]{cancel}

\newcommand{\project}[1]{\textsl{#1}}
\usepackage{xspace}

\newcommand*{\NICER}{\project{NICER}\xspace}
\newcommand*{\MultiNest}{\textsc{MultiNest}\xspace}

\newcommand{\msol}{M$_\odot$~}

\newcommand{\TT}[1]{\texttt{#1}}

\received{August 7, 2019}
\revised{September 6, 2019}
\accepted{September 16, 2019}

\shorttitle{A \NICER VIEW OF PSR~J0030$+$0451: DENSE MATTER}
\shortauthors{Raaijmakers et al.}

\begin{document}

\title{A \NICER VIEW OF PSR~J0030$+$0451: IMPLICATIONS FOR THE DENSE MATTER EQUATION OF STATE}

\correspondingauthor{G.~Raaijmakers}
\email{G.Raaijmakers@uva.nl}

\author[0000-0002-9397-786X]{G.~Raaijmakers}
\author[0000-0001-9313-0493]{T.~E.~Riley}
\author[0000-0002-1009-2354]{A.~L.~Watts}
\affil{Anton Pannekoek Institute for Astronomy, University of Amsterdam, Science Park 904, 1090GE Amsterdam, The Netherlands}

\author{S.~K.~Greif}
\affil{Institut f\"ur Kernphysik, Technische Universit\"at Darmstadt, 64289 Darmstadt, Germany}
\affil{ExtreMe Matter Institute EMMI, GSI Helmholtzzentrum f\"ur Schwerionenforschung GmbH, 64291 Darmstadt, Germany}

\author[0000-0003-4357-0575]{S.~M.~Morsink}
\affil{Department of Physics, University of Alberta, 4-183 CCIS, Edmonton, AB, T6G 2E1, Canada}

\author{K.~Hebeler}
\affil{Institut f\"ur Kernphysik, Technische Universit\"at Darmstadt, 64289 Darmstadt, Germany}
\affil{ExtreMe Matter Institute EMMI, GSI Helmholtzzentrum f\"ur Schwerionenforschung GmbH, 64291 Darmstadt, Germany}

\author{A.~Schwenk}
\affil{Institut f\"ur Kernphysik, Technische Universit\"at Darmstadt, 64289 Darmstadt, Germany}
\affil{ExtreMe Matter Institute EMMI, GSI Helmholtzzentrum f\"ur Schwerionenforschung GmbH, 64291 Darmstadt, Germany}
\affil{Max-Planck-Institut f\"ur Kernphysik, Saupfercheckweg 1, 69117 Heidelberg, Germany}

\author[0000-0002-3394-6105]{T.~Hinderer}
\author[0000-0001-6573-7773]{S.~Nissanke}
\affil{Anton Pannekoek Institute for Astronomy, University of Amsterdam, Science Park 904, 1090GE Amsterdam, The Netherlands}
\affil{GRAPPA Institute of High-Energy Physics, University of Amsterdam, Science Park 904, 1098 XH Amsterdam, The Netherlands}

\author[0000-0002-6449-106X]{S.~Guillot}
\affil{IRAP, CNRS, 9 avenue du Colonel Roche, BP 44346, F-31028 Toulouse Cedex 4, France}
\affil{Universit\'{e} de Toulouse, CNES, UPS-OMP, F-31028 Toulouse, France.}

\author{Z.~Arzoumanian}
\affil{X-Ray Astrophysics Laboratory, NASA Goddard Space Flight Center, Code 662, Greenbelt, MD 20771, USA}

\author[0000-0002-9870-2742]{S.~Bogdanov}, 
\affil{Columbia Astrophysics Laboratory, Columbia University, 550 West 120th Street, New York, NY 10027, USA}

\author[0000-0001-8804-8946]{D.~Chakrabarty}
\affil{MIT Kavli Institute for Astrophysics and Space Research, Massachusetts Institute of Technology, Cambridge, MA 02139, USA}

\author{K.~C.~Gendreau}
\affil{X-Ray Astrophysics Laboratory, NASA Goddard Space Flight Center, Code 662, Greenbelt, MD 20771, USA}

\author[0000-0002-6089-6836]{W.~C.~G.~Ho}
\affil{Department of Physics and Astronomy, Haverford College, 370 Lancaster Avenue, Haverford, PA 19041, USA}
\affil{Mathematical Sciences, Physics and Astronomy, and STAG Research
Centre, University of Southampton, Southampton, SO17 1BJ, UK}

\author{J.~M.~Lattimer}
\affil{Department of Physics and Astronomy, Stony Brook University, Stony Brook, NY 11794-3800, USA}

\author[0000-0002-8961-939X]{R.~M.~Ludlam}
\altaffiliation{Einstein Fellow}
\affiliation{Cahill Center for Astronomy and Astrophysics, California Institute of Technology, Pasadena, CA 91125, USA}

\author{M.~T.~Wolff}
\affil{Space Science Division, U.S. Naval Research Laboratory, Washington, DC 20375-5352, USA}

\begin{abstract}
Both the mass and radius of the millisecond pulsar PSR~J0030$+$0451 have been inferred via pulse-profile modeling of X-ray data obtained by NASA's \NICER mission. In this Letter we study the implications of the mass-radius inference reported for this source by \citet{Riley19} for the dense matter equation of state (EOS), in the context of prior information from nuclear physics at low densities. Using a Bayesian framework we infer central densities and EOS properties for two choices of high-density extensions: a piecewise-polytropic model and a model based on assumptions of the speed of sound in dense matter. Around nuclear saturation density these extensions are matched to an EOS uncertainty band obtained from calculations based on chiral effective field theory interactions, which provide a realistic description of atomic nuclei as well as empirical nuclear matter properties within uncertainties. We further constrain EOS expectations with input from the current highest measured pulsar mass; together, these constraints offer a narrow Bayesian prior informed by theory as well as laboratory and astrophysical measurements. The NICER mass-radius likelihood function derived by \citet{Riley19} using pulse-profile modeling is consistent with the highest-density region of this prior. The present relatively large uncertainties on mass and radius for PSR J0030+0451 offer, however, only a weak posterior information gain over the prior. We explore the sensitivity to the inferred geometry of the heated regions that give rise to the pulsed emission, and find a small increase in posterior gain for an alternative (but less preferred) model. Lastly, we investigate the hypothetical scenario of increasing the \NICER exposure time for PSR~J0030$+$0451. 
\end{abstract}


\section{Introduction} \label{sec:intro}

The cores of neutron stars (NSs) provide a unique environment for exploring matter at densities above nuclear saturation density ($\rho_s=2.7 \times 10^{14}$ g cm$^{-3}$). Theoretical predictions in this regime are diverse, ranging from nucleonic matter under extreme neutron-rich conditions, to stable states of strange matter such as hyperons or deconfined quarks, color superconducting phases, and Bose-Einstein condensates \citep[for recent reviews see][]{Hebeler15,Lattimer16,Oertel17,Baym18}. Our uncertainty about the nature of cold supranuclear-density matter is often encoded in the equation of state (EOS) through general parametric extensions to high densities with an associated prior distribution. Each EOS maps via the stellar structure equations to sequences of stable spacetime solutions given interior boundary conditions \citep[see the review by][]{Paschalidis17}. Properties such as total (or gravitational) mass $M$ and equatorial radius $R_\mathrm{eq}$ of the NS surface feature strongly in the exterior spacetime solution.\footnote{Both in terms of metric functions, and the spatial domain of those functions.} Observational phenomena that are sensitive to the structure of the exterior spacetime, such as the propagation of radiation from the stellar surface to a distant observer, can thus be used to probe the EOS and hence the microphysics of dense matter.

NASA's \textsl{Neutron Star Interior Composition Explorer} \citep[\NICER;][]{Gendreau16}, a soft X-ray telescope installed on the \textsl{International Space Station} in 2017, was developed to estimate masses and radii of NSs using pulse-profile modeling of nearby rotation-powered millisecond pulsars (MSPs). The magnetic polar caps of MSPs, thought to be heated by (return) currents in the pulsar magnetosphere, produce thermal emission in the soft X-ray band \citep{Harding02}. As the MSP rotates, this emission gives rise to perceived pulsations, and relativistic effects encode information about the spacetime into the phase-energy resolved pulse-profile.\footnote{A pulse-profile consists of X-ray counts per rotational phase bin per instrument detector channel, curated by phase-folding X-ray events according to a pulsar timing ephemeris.} Pulse-profile modeling employs relativistic ray-tracing and Bayesian inference software to jointly infer mass and radius \citep[see][for an overview of the technique]{Bogdanov16b,Watts16,Watts19b}. 

\citet{Riley19} jointly estimated the mass $M$ and radius $R_{\rm eq}$ of the MSP PSR~J0030$+$0451 conditional on \NICER X-ray Timing Instrument (XTI) photon event data curated by \citet{Bogdanov19a}. The results derived are also conditional upon the modeling choices made in the analysis, e.g.,: the assumption of two disjoint surface hot regions, each with some local comoving effective temperature field but no magnetic field physics; a fully ionized hydrogen atmosphere; and a specific parameterization of the uncertainty in the \NICER XTI instrument response.\footnote{For an independent analysis of the same data set using different modeling choices and methodology see \citet{Miller19b}, which follows the approach outlined in \citet{Miller2019}.} The restriction to two disjoint hot regions was motivated by the presence of two distinct pulses in the observed pulse profile. \citet{Riley19} allowed for the possibility of the hot regions being non-antipodal and non-identical, and considered various shapes for the hot regions including circles, rings (with the centers both concentric and offset) and crescents, filled with material of a single local comoving temperature. Model comparison enabled the identification of a favored configuration, using a combination of performance measures including the evidence (the prior predictive probability of the data) and graphical posterior predictive checking (to verify whether or not an updated model generates synthetic data\footnote{For illustration, a pulse-profile count-number data set can be simulated given specific instances of the following components: a spacetime solution; a surface hot-region configuration (effective temperature, geometry); an atmospheric beaming function (composition, ionization); background contribution (astrophysical, instrumental); an instrument response function; and a noise model (Poissonian). Source emission is propagated via relativistic ray-tracing through the spacetime towards a distant observer inclined to the stellar spin axis, and is subsequently operated on by the instrument response function; the product is a joint sampling distribution for photon count numbers, which is intrinsic to the definition of a likelihood function. The notion of synthetic data generation is a vital part of the Bayesian inference framework. See \citet{Riley19} and \citet{Watts19b} for more discussion.} without obvious residual systematic structure in comparison to the real data). 

For the family of models considered, the favored configuration is one in which the hot regions consist of a small hot spot with angular extent of only a few degrees, and a more extended hot crescent, both in the same rotational hemisphere (referred to in \citealt{Riley19} as \texttt{ST+PST}). For this configuration, the inferred mass and equatorial radius\footnote{With respect to a Schwarzschild coordinate chart, see section 2.3.1 of \citet{Riley19} for more details.} are $M=1.34^{+0.15}_{-0.16}$ \msol and $R_\mathrm{eq} = 12.71^{+1.14}_{-1.19}$ km. The compactness $GM/R_\mathrm{eq}c^2 = 0.156^{+ 0.008}_{-0.010}$ is more tightly constrained.

The credible bounds reported here are approximately the $16\%$ and $84\%$ quantiles in marginal posterior mass. The spin frequency of PSR~J0030$+$0451 is only 205~Hz: $M$ and $R_{\rm eq}$ can, due to the size of the credible intervals, therefore be identified as those of a non-rotating star with an equivalent number of baryons. The effects of rotation are discussed in more detail in Section~\ref{discussion}.

If the extended hot region is restricted to have ring-like topology rather than that of a simply connected crescent \citep[\texttt{ST+CST} in][]{Riley19}, the inferred mass and equatorial radius are $M=1.44^{+0.18}_{-0.19}$ \msol and $R_\mathrm{eq} = 13.89^{+1.22}_{-1.39}$ km. The compactness $GM/R_\mathrm{eq}c^2 = 0.16 \pm 0.01$ is however shared with \TT{ST+PST}---at the quoted precision. Although \TT{ST+CST} was not the favored configuration \textit{a posteriori}, it provides a useful illustration of the sensitivity of dense matter inferences to the nuisance parameters controlling the surface radiation field. As  pulsar theory develops, dense matter inferences therefore need to be re-examined in step. Fortunately, such calculations are less expensive to execute given posterior samples because nuisance-parameter marginalization is thereby approximated.

In this Letter, we examine how the constraints on NS mass and radius translate into constraints on the dense matter EOS. Ultimately, we intend to carry out a population-level analysis conditional on all \NICER MSP targets, in order to report a joint summary for \NICER. We propose to inject as little information as is reasonable from other statistical constraints derived from astronomical data sets---the exception being information from a radio pulsar mass measurement. Eventually, we aim to combine the joint \NICER constraints with those derived using other missions, where appropriate. However, for now we have information for a single source, and in this Letter we address how the joint mass-radius information derived by \citet{Riley19} for PSR~J0030$+$0451 maps to constraints on the dense matter EOS. The second principal aim of this Letter is therefore to formalize a plan for post-processing posterior information derived via pulse-profile modeling, into posterior information about dense matter. The post-processing phase for dense matter study is far less computationally expensive than the preceding X-ray analysis in which the likelihood information relevant for dense matter study is computed. We can therefore effectively update our posterior information on-the-fly as new information becomes available, by jointly compiling \NICER source-by-source nuisance-marginalized likelihood functions into posterior constraints about a common EOS.

\section{\NICER EOS constraints}

\subsection{EOS parameterizations}
Following the methods described in \citet{Greif19}, we model the interior of PSR~J0030$+$0451 using two distinct EOS parameterizations: the piecewise-polytropic (PP) model from \citet{Hebeler13} and a speed of sound (CS) model introduced in \citet{Greif19}---see also \citet[][]{Tews18b}. These parameterizations were matched at $1.1~\rho_s$ to either the upper limit or the lower limit of a calculated EOS range based on chiral effective field theory (cEFT) interactions including theoretical uncertainties \citep[for details, see][]{Hebeler10a, Hebeler13}. This discrete matching leads to a bimodality in the prior of the EOS, which we mitigate here by introducing an additional parameterization of the EOS inside the cEFT band. For simplicity we assume a single polytrope, i.e.,
\begin{align}
    P(\rho) = K \left(\frac{\rho}{\rho_{s}}\right)^{\Gamma},
    \label{eq:polytrope}
\end{align}
where $P$ is the pressure, $\rho$ the baryon mass density, $K$ (in units of MeV fm$^{-3}$) a free parameter, and $\Gamma$ the adiabatic exponent. We determine $\Gamma$ and the bounds of $K$ by fitting Equation~\eqref{eq:polytrope} to the lower and upper limits of the cEFT band for densities between 0.5$\rho_s$ and 1.1 $\rho_s$ and find that these limits are well approximated by $\Gamma = 2.5$, $K_{\mathrm{min}} = 1.70$, and $K_{\mathrm{max}} = 2.76$. At densities below 0.5 $\rho_s$ we match to a single crust EOS \citep[BPS; see][]{Baym71}. Comparing the full range of masses and radii permitted by EOS under both parameterizations with a continuous matching to the the cEFT band against the upper/lower limit case, we find they are consistent, although a slightly larger range is obtained for the PP model for the continuous case. This is due to the polytropic fit to the cEFT band allowing for a small set of additional EOS that are soft enough in the low-density regime to result in small NS radii but stiff enough at  larger densities to comply with the pulsar mass constraint (see Sec. \ref{subsec:priors} and \citet{Greif19}).

\begin{figure*}[t!]
\centering
\includegraphics[width=0.9\textwidth]{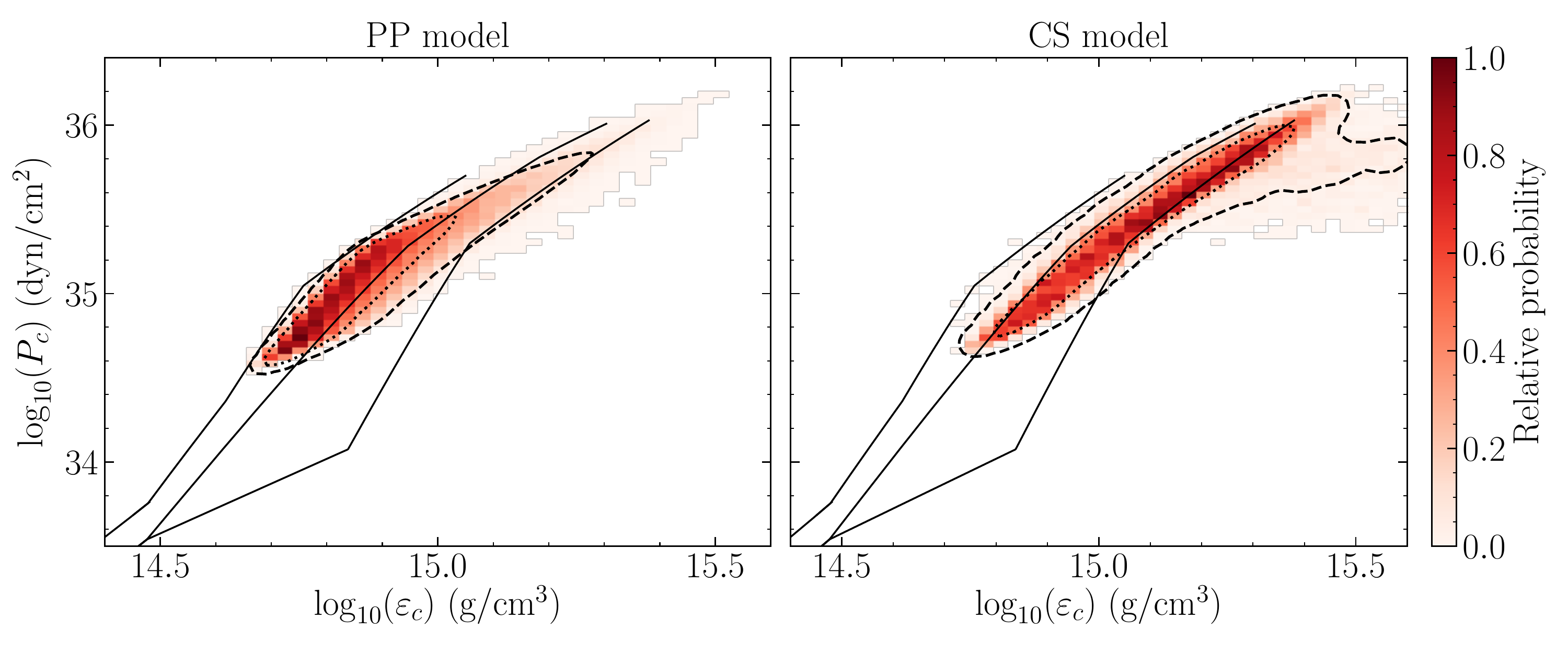}
\includegraphics[width=0.9\textwidth]{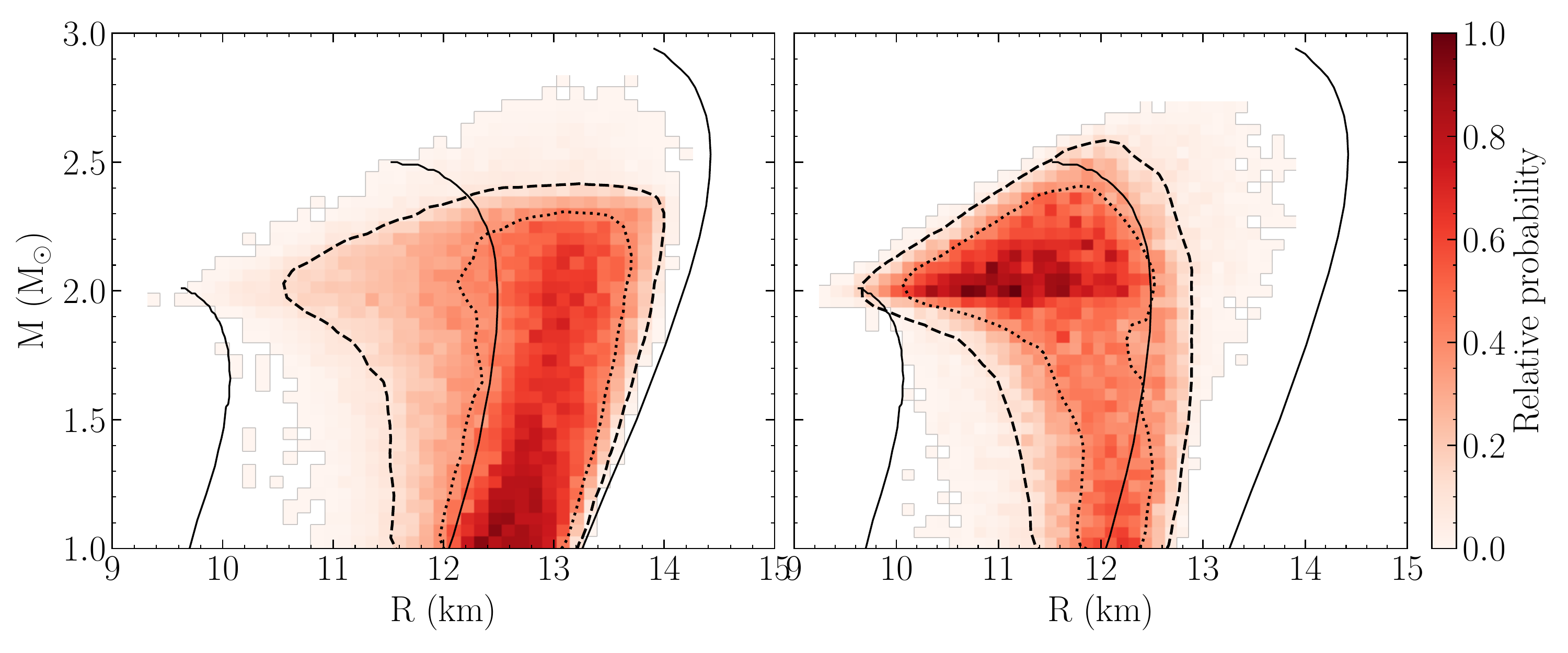}
\caption{Prior probability distributions for PSR~J0030$+$0451 transformed to the joint space of central pressure and central energy density (upper panels), and the space of mass and radius (lower panels), for the PP (left panels) and CS parameterization (right panels). The dotted and dashed contours bound the highest-density two-dimensional regions respectively containing $68\%$ and $95\%$ of the prior mass. The peak in the CS model just above $1.97$~M$_{\odot}$ is due to how the model is constructed: all EOS are forced to soften at high densities to comply with causality and at asymptotic densities with the constraint from pQCD, causing all mass-radius sequences to have $\partial M/\partial R \lesssim 0$ near the maximum mass, thereby overlapping each other \citep[see][]{Greif19}. Note that these priors are specifically for PSR~J0030$+$0451 because adjustments are made to match the priors in the analysis in \citet{Riley19}; i.e., $M\in [1.0, 3.0]$~M$_{\odot}$ and $R\in[3 r_g, 16]$~km where $r_g(M) = GM/c^2$. For comparison the three representative EOS from \citet{Hebeler13} are shown as solid curves: HLPS Soft, Intermediate, and Stiff. Note that the discernible small-scale structure is due to: (i) the behavior of the (numerical) transformation from interior matter parameters to exterior spacetime parameters; and (ii) finite sampling noise.}
\label{fig:fig1}
\end{figure*}

\subsection{Bayesian framework}
\label{subsec:bayesian}

To derive constraints on the EOS from a single-star mass-radius posterior density distribution, we use the Bayesian framework outlined in \citet{Greif19} and \citet{Riley18}. Let us combine the EOS parameters and the central density $\varepsilon_{c}$ of PSR~J0030$+$0451 into a vector $\bm{\theta}$. The posterior distribution of $\bm{\theta}$ is proportional to the product of the prior distribution of $\bm{\theta}$ and the \textit{nuisance-marginalized} likelihood function of $\bm{\theta}$ (Bayes' theorem):
\begin{equation}
\label{eq:eq1}
\begin{split}
p(\bm{\theta} \,|\, \bm{d}, \mathcal{M})
&
\propto 
p(\bm{\theta} \,|\, \mathcal{M})
~
p(\bm{d} \,|\, \bm{\theta}, \mathcal{M}) \\
&
\propto
p(\bm{\theta} \,|\, \mathcal{M})
~ 
p(M, R \,|\, \bm{d}, \mathcal{M}),
\end{split}
\end{equation}
where $\bm{d}$ denotes the \NICER PSR~J0030$+$0451 data set, and $\mathcal{M}$ denotes the model. The model includes the physics of processes both interior and exterior to the star: the EOS and central conditions (present work); and X-ray emission, propagation, and detection \citep[][and references therein]{Riley19}. Implicitly, the model includes all Bayesian prior information. The parameters $\bm{\theta}$ map deterministically to the mass $M=M(\bm{\theta};\Omega)$ and radius $R=R(\bm{\theta};\Omega)$, where the coordinate angular rotation frequency $\Omega=0$ (see Section~\ref{subsec:rotation}). All parameters apart from $M$ and $R$ are, for the purposes of the discussion that follows, termed {\it nuisance parameters} and are marginalized out.\footnote{Note that they do, however, describe important and interesting physics on the surface of the star and exterior to it, to which our pulse-profile modeling is extremely sensitive. See for example \citet{Bilous19} on the implications of some of these inferred ``nuisance parameters'' for our understanding of pulsar magnetospheres.}

To obtain the second line of Equation~(\ref{eq:eq1}) we equated the nuisance-marginalized likelihood function of $M$ and $R$ to the nuisance-marginalized joint posterior density distribution of $M$ and $R$ reported by \citet{Riley19}. This proportionality holds exactly because the marginal joint prior distribution of $M$ and $R$ chosen by \citet{Riley19} is jointly flat. Our numerical nuisance-marginalized likelihood function is an approximation to the exact nuisance-marginalized likelihood function because we are post-processing posterior samples, and because post-processing involves kernel density estimation (KDE) of the posterior density function. We then sample from the posterior density $p(\bm{\theta}\,|\,\bm{d},\mathcal{M})$ in Equation~(\ref{eq:eq1}) using the nested sampling software \MultiNest \citep{Feroz08, Feroz09, Feroz13, Buchner14}.

\subsubsection{Priors}
\label{subsec:priors}
The prior density, $p(\bm{\theta}\,|\,\mathcal{M})$, in Equation~(\ref{eq:eq1}) is identical to the prior described in Section~3.1.1 of \citet{Greif19}, with the exception of the implementation of the continuous range within the cEFT band. Here we introduce an additional uniform prior on the parameter $K$ in Equation~(\ref{eq:polytrope}) with support $K\in[1.70, 2.76]$. The prior for all parameters in the PP and CS model is summarized in Table \ref{table:tab1}. Moreover, we require that every EOS assigned a finite local prior density can support a stable $1.97$~M$_{\odot}$ NS, equal to the lower $1\sigma$ limit on the mass of the most massive NS measured to date \citep[PSR~J0348$+$0432;][]{Antoniadis13}.\footnote{Note that this is an ad-hoc interpretation of the information encoded in the pulsar mass measurement. We discuss this in more detail in Section~\ref{subsec:discuss prior implementation}.}

The joint prior density $p(\bm{\theta}\,|\,\mathcal{M})$ is written (for both parameterizations) as
\begin{equation}
p(\bm{\theta}\,|\,\mathcal{M})
=
p({\varepsilon_{c}\,|\,{\rm EOS},\mathcal{M}})
p({\rm EOS}\,|\,\mathcal{M}),
\end{equation}
where the conditional prior density of the central density $\varepsilon_{c}$, $p(\varepsilon_{c}\,|\,\rm{EOS},\mathcal{M})$, is numerically evaluated to impose global spacetime stability. Given an EOS, we inverse sample the conditional density with rejection: we reject $\bm{\theta}$ if $\bm{\theta}\mapsto(M,R)$ does not yield a stable spacetime solution that exists within the support of the \textit{posterior} density $p(M,R\,|\,\bm{d},\mathcal{M})$. Outside of this support, the nuisance-marginalized likelihood function has not been estimated in the preceding X-ray analysis. \citet{Riley19} imposed prior support with hard bounds of $M\in [1.0, 3.0]$~M$_{\odot}$ and $R\in[3 r_g, 16]$~km, where $r_g = GM/c^2$. Note that, except for the lower bound on the mass, these bounds already have zero support from the prior on the EOS model. 
The lower bound on the mass is implemented in \MultiNest by assigning a likelihood value below a certain threshold to any mass-radius pair outside this bound. Any mass-radius pair with a likelihood lower than this threshold will then be ignored in the nested sampling process by \MultiNest (see T. E. Riley \& A. L. Watts, in preparation, for a discussion on prior density implementation options for use with \MultiNest). Besides the matching to the cEFT band, the PP model is constrained by causality and the requirement to support a $1.97$\,M$_\odot$ NS. Thus, the PP model has prior support for the central density up to the maximum mass (which is required to be reached before the speed of light $c_s = c$).
For the CS model we also consider the EOS for central densities up to the maximum mass and impose the constraints that
\begin{enumerate}[(i)]
\item the speed of sound for all energy densities is less than the speed of light; 
\item the speed of sound of each EOS converges  to $(c_{s}/c)^2 = 1/3$ from below at $\sim\!50 \rho_s$, following the calculations of the speed of sound for asymptotically high densities by perturbative quantum chromodynamics (pQCD; \citet{Fraga14});
\item the bulk properties of matter at densities $\rho \le 1.5 \rho_s$ can be described as a normal Fermi liquid, which restricts the speed of sound at these densities to be $(c_{s}/c)^{2} \le 0.163$ \citep[see][and references therein, for more detail]{Greif19}.
\end{enumerate}

While both our prior for the CS model and the PP model allow for phase transitions at certain density ranges, they do not cover all possibilities of transitions to other forms of matter. However, EOS that mimic hybrid stars for which the transition is smooth do exist within our prior bounds. We discard any EOS model that allows for two disconnected stable branches on the mass-radius sequence \citep[similar to][]{Alford16}, which occurred in our sampling for certain parameter sets of the PP model but is also possible in the CS model.

\begin{figure*}[t!]
\centering
\includegraphics[width=0.9\textwidth]{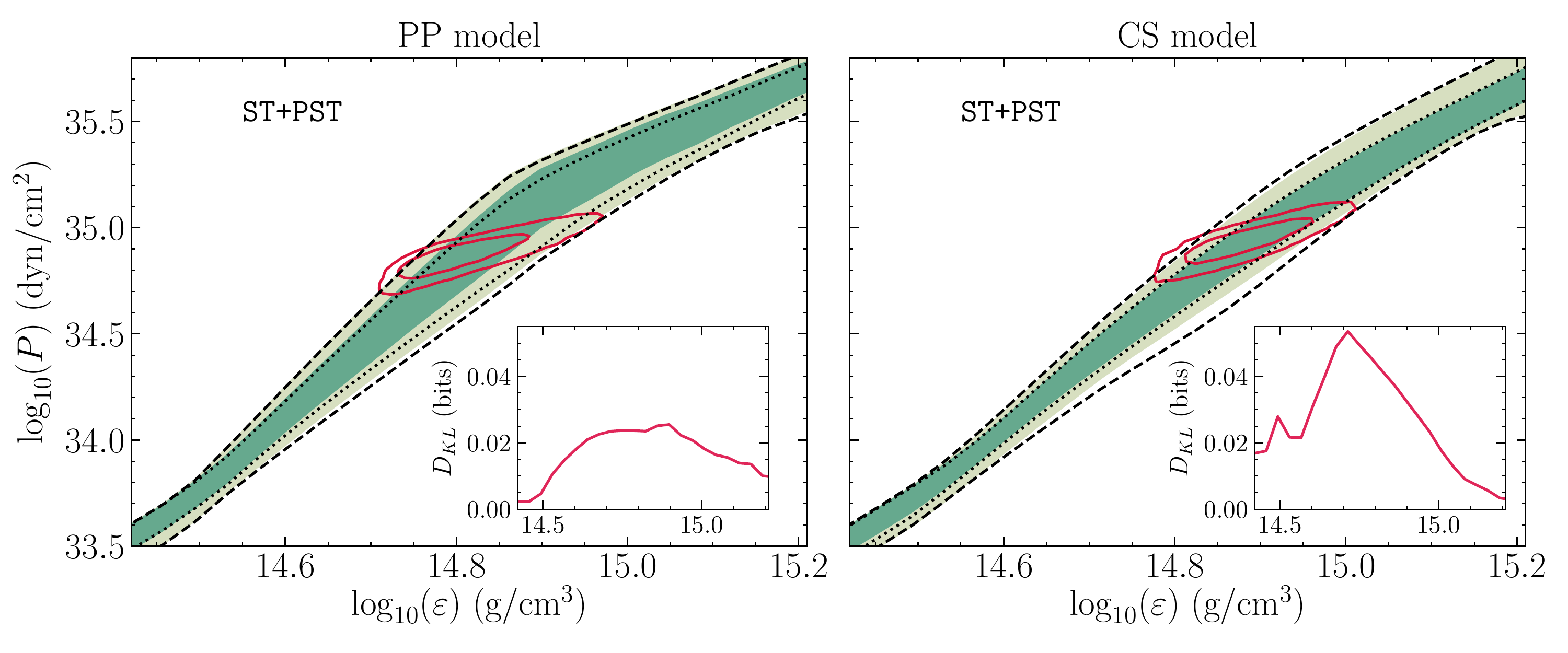}
\includegraphics[width=0.9\textwidth]{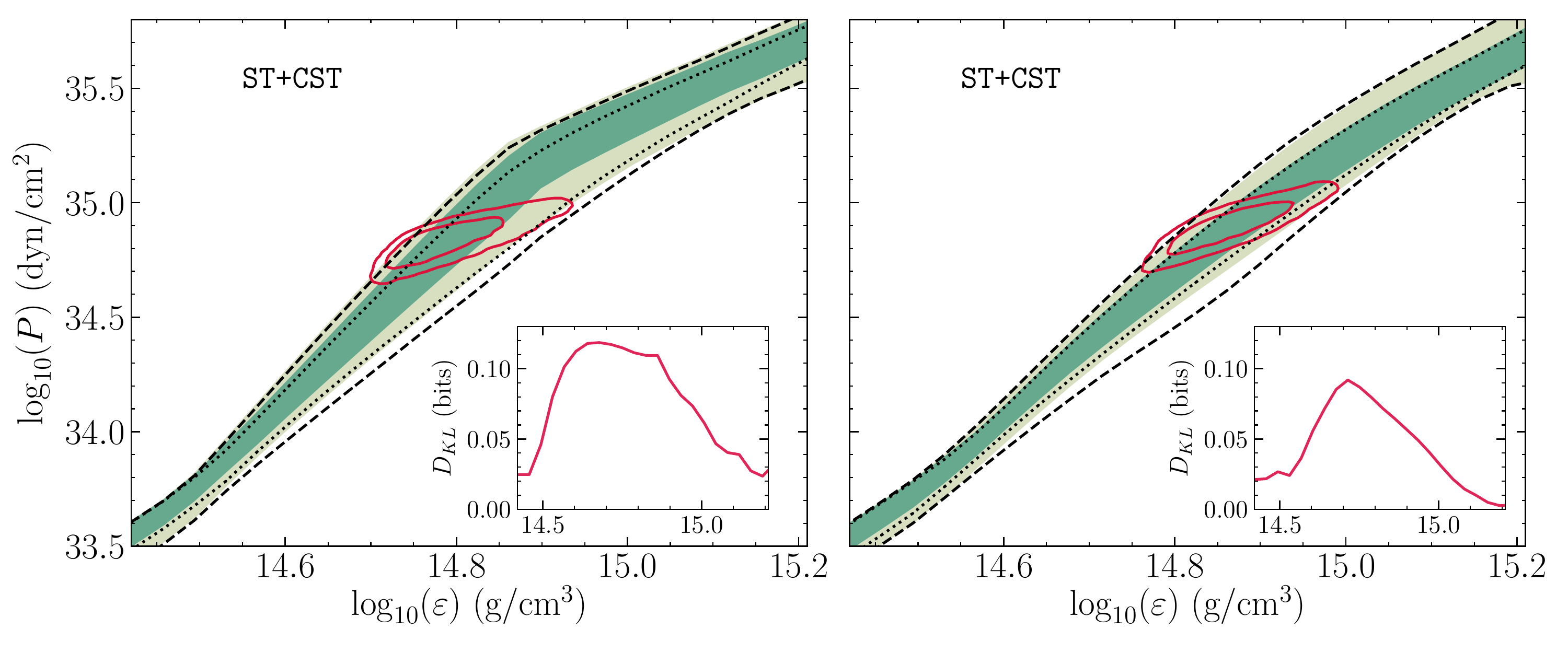}
\caption{Marginal posterior distributions of the pressure $P$ conditional on energy density $\varepsilon$, $p(P\,|\,\varepsilon,\bm{d},\mathcal{M})$, for the PP model (left) and the CS model (right), and for both the preferred \texttt{ST+PST} model (upper panels) and the alternative \texttt{ST+CST} model (lower panels). At each value of $\varepsilon$, there exist $68\%$ and $95\%$ posterior credible intervals for the pressure $P$; we connect these intervals to form the shaded bands. The black dotted and dashed lines respectively indicate the joined $68\%$ and $95\%$ credible interval bands, but for the conditional and marginal prior distribution, $p(P\,|\,\varepsilon,\mathcal{M})$. The red contours in each panel indicate the $68\%$ and $95\%$ highest-density posterior credible regions of central energy density and central pressure. Constraints on the EOS for densities higher than these contours are only determined by our choice of parameterization and are not directly informed by the mass-radius likelihood  function (and thus in turn, the data). The lower-right inset panels illustrate the evolution of the Kullback-Leibler (KL) divergence with respect to the energy density, showing that most posterior information is gained for densities below $10^{15}$~g cm$^{-3}$, not coincidentally the highest possible central density reached in PSR~J0030$+$0451. Note that due to finite sampling noise the precise features of the evolution of the KL divergence might be disputed, but the global trend of the curve is unaffected.}
\label{fig:fig2}
\end{figure*}

In order to understand our prior choices and the effect of the continuous matching to the cEFT band in more detail, we randomly sample $\sim\!10^5$ points from the prior distributions and calculate for each point: (i) a central energy density and central pressure pair; and (ii) a mass-radius pair. The resulting distributions for both the PP and CS model are shown in Figure~\ref{fig:fig1}. We note that the bimodality observed in the prior in \citet{Greif19} has been smoothed out by the addition of the continuous matching. The darker region at low density, or equivalently at low masses, comes as a result of all EOS being matched to the cEFT band. However, in the analysis of PSR~J0030$+$0451 this region is outside of the prior support ($M\geqslant 1.0$~M$_{\odot}$) from the analysis in \citet{Riley19}. Less intuitive is the darker region for the prior of the CS model just above $1.97$~M$_{\odot}$, which can be explained by investigating the individual mass-radius sequences that contribute to this clustering: the CS model is constructed to be causal at all densities and is only truncated when $dM/d\varepsilon_c \leq 0$. This causes the mass-radius sequences to bend over on the $M$-$R$ plot after the $1.97$~M$_{\odot}$ constraint is fulfilled and extend horizontally toward smaller radii, overlapping with each other. The PP model differs in this feature because individual EOS are truncated when the EOS reaches $c_s = c$, so that such an EOS does not need to bend over. This allows mass-radius sequences with steep slopes to exist up to the density where $c_s = c$.

\subsection{EOS constraints Based on PSR~J0030$+$0451}
We consider two distinct mass-radius posterior distributions supplied by \citet{Riley19}, each conditional on assumptions about the thermally emitting hot regions on the surface. These assumption sets are identified as the \texttt{ST+PST} and \texttt{ST+CST} models, which yielded $68\%$ credible intervals on the equatorial radius of $R = 12.71^{+1.14}_{-1.19}$ km and $R = 13.89^{+1.23}_{-1.38}$ km, respectively.  We stress that the favored configuration identified by \citet{Riley19} is \texttt{ST+PST}; we explore the other configuration only to illustrate the potential sensitivity of our conclusions to developments in pulsar theory or additional observations.

We show the $68\%$ and $95\%$ credible regions of the resulting posterior distributions transformed to the EOS space in Figure~\ref{fig:fig2}. Sensitivity to compactness manifests strongly as a constraint on central conditions---the density and pressure. Note that the marginal posterior credible interval on the central density of PSR~J0030$+$0451 is dependent (or conditional) on the assumed EOS model. The inset panels in Figure~\ref{fig:fig2} show that most information gain about matter pressure, measured through the Kullback-Leibler divergence\citep[][see below]{kullback1951}, is in the vicinity of densities found in the core, but in absolute terms the gain is negligible. The EOS at higher densities than the central density (see also Table \ref{table:tab2}) is not directly informed by the data because \textit{a posteriori}, matter at such densities does not exist in the star; any information gain is due to dependence on our choice of EOS parameterization which couples low- and high-density regimes via simple functional forms. Also rendered is a comparison of the posterior distributions with the prior distributions (compare the green shaded bands and the black dashed and dotted bands), which suggests no remarkable reduction in the degree of uncertainty---the prior distribution of the EOS functions is dominant \citep[just as in][]{Greif19}. In addition, the sensitivity of the posterior distributions to the chosen EOS parameterization (PP or CS) is stronger than to the model chosen for the hot regions (\texttt{ST+PST} or \texttt{ST+CST}). This can also be concluded from the inferred values of central energy densities and corresponding pressures in Table \ref{table:tab2}; larger differences occur between chosen parameterizations than models for hot regions. As a consequence it is difficult to investigate the sensitivity of our inference to the assumed hot region model.

\begin{table}[t!]
\centering
\begin{tabular}{@{}llll@{}}
\toprule
Parameterization & \multicolumn{1}{p{2cm}}{\raggedright Hot Region \\ Model} & $\log_{10}(\varepsilon_c)$ & $\log_{10}(P_c)$ \\
\midrule
PP model & \texttt{ST+PST} & 14.80$^{+0.04}_{-0.07}$ &  34.87$^{+0.06}_{-0.08}$\\
 & \texttt{ST+CST} & 14.78$^{+0.04}_{-0.06}$ &  34.84$^{+0.06}_{-0.09}$ \\
 CS model & \texttt{ST+PST} & 14.88$^{+0.05}_{-0.06}$ &  34.94$^{+0.07}_{-0.09}$ \\
 & \texttt{ST+CST} & 14.86$^{+0.05}_{-0.06}$ &  34.90$^{+0.07}_{-0.1}$ \\
 \bottomrule
 \midrule
\end{tabular}
\caption{Inferred median values of the central energy density and corresponding central pressure for PSR J0030+0451, where the errors indicate the interval that contains $68\%$ of the posterior mass. }
\label{table:tab2}
\end{table}

\begin{figure*}[t!]
\centering
\includegraphics[width=0.9\textwidth]{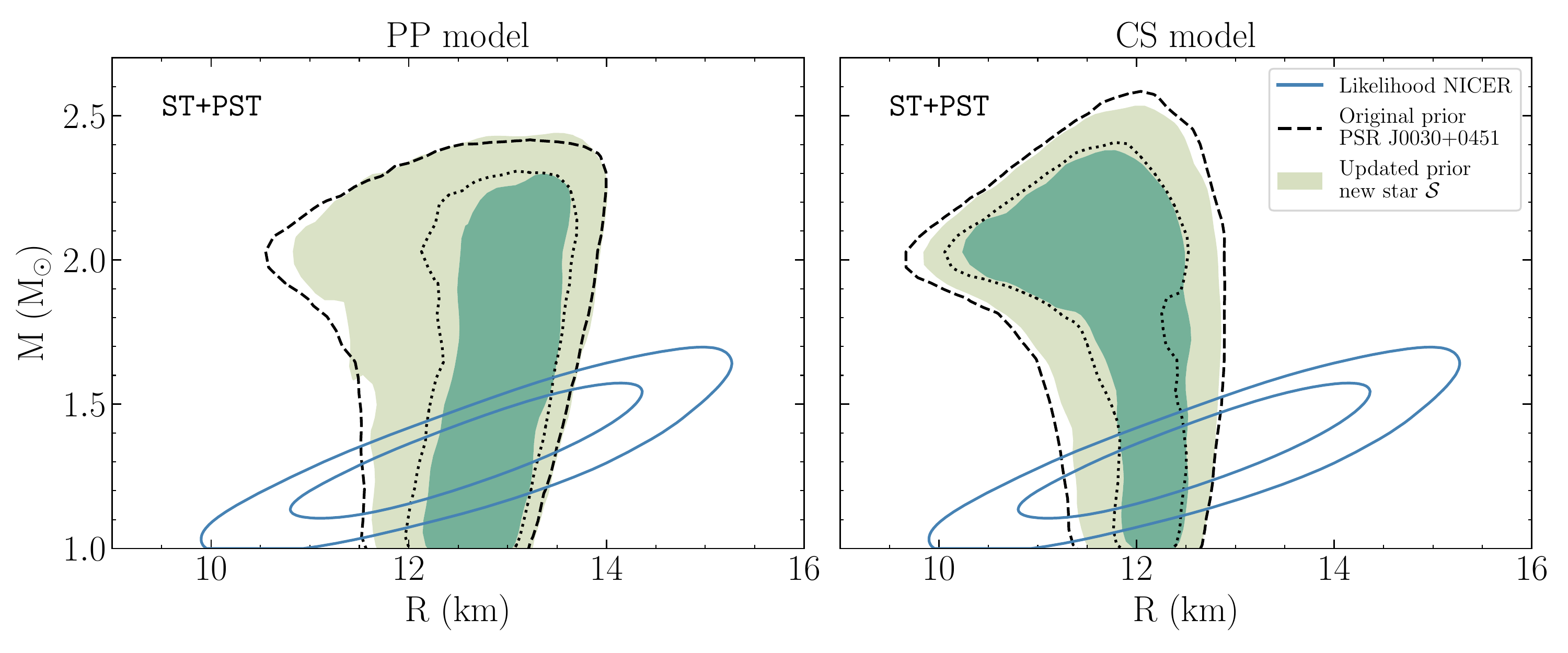}
\includegraphics[width=0.9\textwidth]{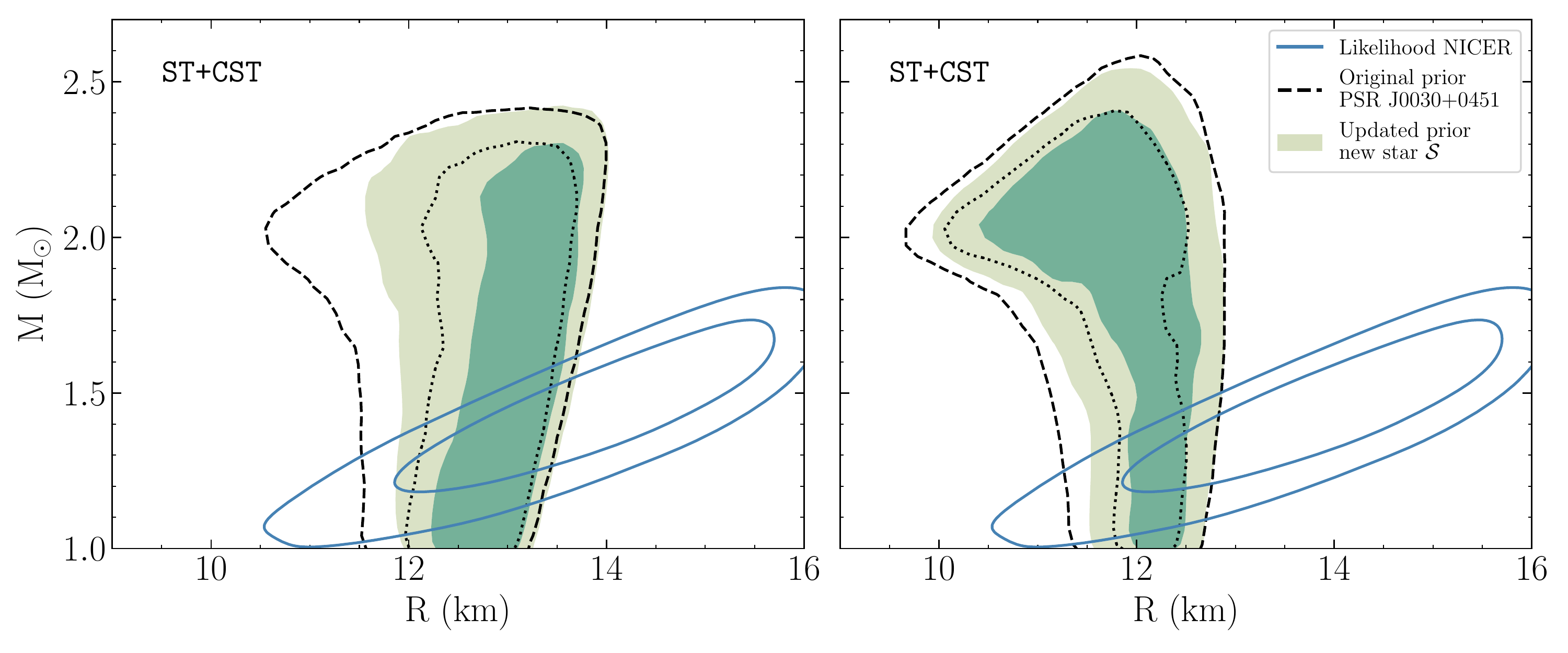}
\caption{The $68\%$ and $95\%$ highest-density credible regions of the PSR~J0030$+$0451 likelihood function (normalized by a flat joint density) are bounded by the blue contours: the \texttt{ST+PST} model is featured in the upper panels, and the \texttt{ST+CST} model is featured in the lower panels. The black contours are identical to those shown in Figure~\ref{fig:fig1}, being associated with a population-level prior transformed to mass-radius space; here we show how this prior is updated as a result of our analysis of PSR~J0030$+$0451. To provide context, we are considering the implications for a hypothetical future analysis of data from a different pulsar $\mathcal{S}$ (i.e., not PSR~J0030$+$0451). We assume that the EOS of all matter (core to crust) is shared between $\mathcal{S}$ and PSR~J0030$+$0451; as a joint prior for the EOS of matter in $\mathcal{S}$, we thus invoke the joint posterior distribution of EOS parameters conditional on \NICER observations of PSR~J0030$+$0451. We assume the central density of $\mathcal{S}$ is drawn from the same population-level density $p(\varepsilon_{c}\;|\;\mathrm{EOS})$ as PSR~J0030$+$0451. We then transform the joint prior of the EOS parameters and central density of $\mathcal{S}$ to the joint space of the mass and radius of $\mathcal{S}$; we render the two-dimensional regions enclosing $68\%$ and $95\%$ of the updated prior mass in green. We note that the updated prior distributions are still mostly dominated by the prior for PSR J0300+0451, with slightly more support for higher radii in the \texttt{ST+CST} model.}
\label{fig:fig3}
\end{figure*}

In Figure~\ref{fig:fig3} we show the nuisance-marginalized likelihood function of the mass and radius of PSR~J0030$+$0451. We also show how the analysis of the dense matter in this source modifies our population-level prior, when transformed to mass-radius space (an alternative representation of the posterior information on the EOS). Consider the hypothetical future analysis of some other observed NS, $\mathcal{S}$, which shares an EOS---from core to crust---with PSR~J0030$+$0451, and whose central density $\varepsilon_c$ is drawn from the same population-level density $p(\varepsilon_c~|~\text{EOS})$ as that of PSR~J0030$+$0451.\footnote{Note that this a Bayesian hierarchical model, where the shared EOS parameters appear in the likelihood function but also effectively appear as hyperparameters of the central density prior distribution.}
The joint prior for analysis of $\mathcal{S}$ is then based on the PSR~J0030$+$0451 posterior (see the caption for details). The figure clearly illustrates that \textit{after} learning about the EOS from PSR~J0030$+$0451, the prior distribution of mass and radius of $\mathcal{S}$ remains dominated by the original prior information invoked for analysis of PSR~J0030$+$0451. Note that considering the \texttt{ST+CST} model for PSR~J0030$+$0451 shifts the prior mass-radius sequences toward slightly higher radii. \footnote{To compute the $68\%$ and $95\%$ highest-density credible regions in Figure \ref{fig:fig3} we have applied a Gaussian kernel density estimation with the bandwidth parameter according to Scott's rule \citep{Scott92}, and determined that the results and conclusion presented here are not affected by this parameter.}

The posterior distributions on the speed of sound for the CS model (not shown in this Letter) are similar to the distributions shown in figure~8 of \citet{Greif19}, again showing no evidence of the speed of sound reaching the asymptotic limit $(c_{s}/c)^2 = 1/3$ within the range of energy densities relevant for NSs.

To quantify the information gain (in bits) of the posterior over the prior, we compute for each model the KL divergence, an asymmetric measure of how different one probability distribution is from another \citep{kullback1951}.\footnote{See Appendix A.2.4 of \citet{Riley19} for supplementary detail about the KL divergence.} 
The errors on the divergences are obtained by repeated calculation for each in a set of posterior realizations; each realization is simply simulated by bootstrap resampling with replacement from the samples according to their importance weights. Note that this is intended as a fast and simple approximation to the nested-sampling error theory treated generally in the literature \citep[namely,][]{skilling2006, higson2018sampling,higson2019diagnostic,Higson_nestcheck}, in order to get a handle on the magnitude of the noise.
 
Inspecting individual KL divergences for each parameter (see Table~\ref{table:tab1}) reveals that most information is gained in the distribution of central densities, while other parameters have KL divergences closer to zero. This can be visualized by comparing the posterior on central energy densities and pressures in Figure~\ref{fig:fig2} with the prior distribution in Figure~\ref{fig:fig1}: the most information is gained along direction of the central energy density. 

We further quantify the posterior distributions by performing a model comparison between the PP and CS model using Bayes' factors. The Bayes' factor is the ratio between the model evidences;\footnote{The model evidence being the principal computational target of the \MultiNest algorithm.} if we were to accept a uniform prior mass distribution over the discrete models, the posterior odds ratios are equal to the Bayes' factors. In Table \ref{table:tab1} we report the evidences as well as the Bayes' factors, computed here as the ratio of the PP model evidence to the CS model evidence for some exterior-physics (\TT{PST} or \TT{CST}) likelihood function. Following the interpretation of Bayes' factors of \citet{KassRaft95}, we observe that there is no preference for either of the two parameterizations when using both the \texttt{ST+PST} and the \texttt{ST+CST} model. The Bayes' factors show however slightly increased support for the PP model when using the \texttt{ST+CST} model, a consequence of the tighter constraint on large radii for the CS model. As a result there is a tension between the inferred radius $R = 13.89^{+1.23}_{-1.38}$ km for the \texttt{ST+CST} model and the allowed range of radii in the CS model, which is less evident in the PP model.

\section{Effect of Increased Exposure Time}\label{sec:simulated obs time ext}

\begin{figure*}[t!]
\centering
\includegraphics[width=0.9\textwidth]{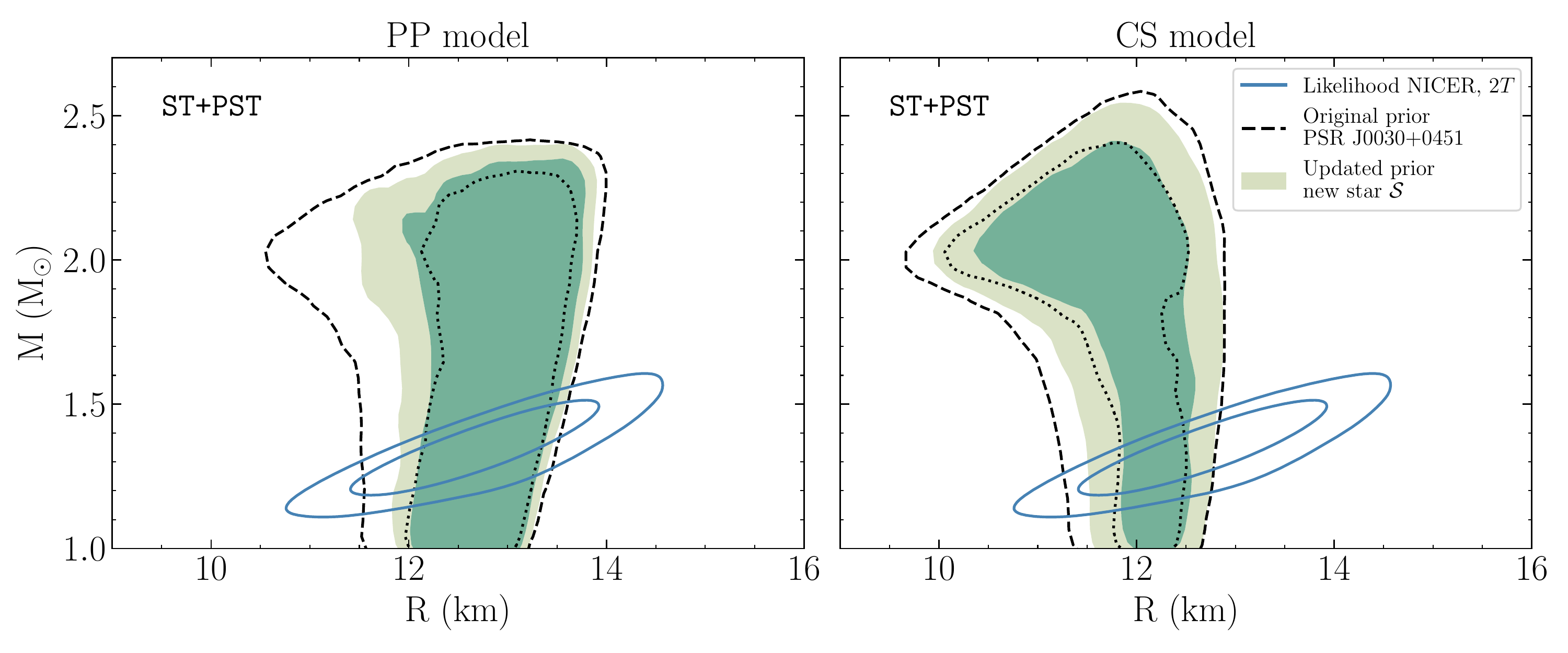}
\includegraphics[width=0.9\textwidth]{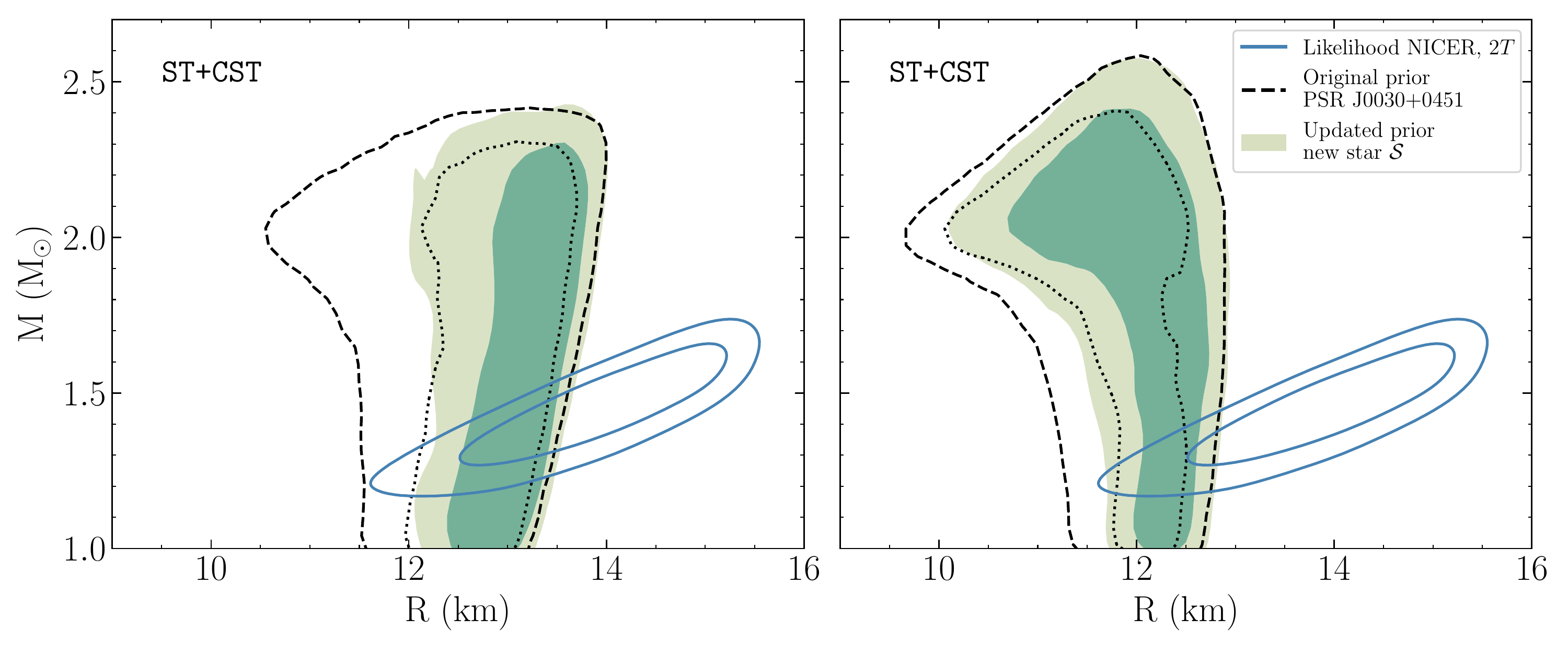}
\caption{Similar to Figure~\ref{fig:fig3} but for the hypothetical scenario where the observing time of PSR~J0030$+$0451 is increased. We crudely simulate the evolution in the nuisance-marginalized likelihood function by contracting the $p(M,R\;|\;\bm{d})$ credible regions by a factor of two, while retaining the coordinates of the point which reports the highest nuisance-marginalized likelihood value.}
\label{fig:fig5}
\end{figure*}

\begin{table*}[t!]
\centering
\begin{tabular}{@{}lllllll@{}}
\toprule
 & Parameter & Prior Density & \multicolumn{1}{c}{\texttt{ST+PST} $(T)$} & \multicolumn{1}{c}{\texttt{ST+CST} $(T)$} & \multicolumn{1}{c}{\texttt{ST+PST} $(2T)$} & \multicolumn{1}{c}{\texttt{ST+CST} $(2T)$} \\ \midrule
 & & & $\widehat{D}_{\rm KL}$~(10$^{-3}$ bits) & $\widehat{D}_{\rm KL}$~(10$^{-3}$ bits) & $\widehat{D}_{\rm KL}$~(10$^{-3}$ bits) & $\widehat{D}_{\rm KL}$~(10$^{-3}$ bits) \\
\midrule
PP model & K & \textit{U}(1.7, 2.76) & 2.47 $\pm$ 0.27 & 26.24 $\pm$ 1.13 & 3.72 $\pm$ 0.37 & 74.80 $\pm$ 7.94   \\
  & $\Gamma_1$ & \textit{U}(1, 4.5) & 16.32 $\pm$ 1.00 & 72.77 $\pm$ 1.89 & 36.83 $\pm$ 1.31 & 129.98 $\pm$ 11.62  \\
  & $\Gamma_2$ & \textit{U}(0, 8) & 3.09 $\pm$ 0.40 & 7.57 $\pm$ 0.61 & 5.27 $\pm$ 0.52 & 6.72 $\pm$ 1.83\\
  & $\Gamma_3$ & \textit{U}(0.5, 8) & 6.64 $\pm$ 0.24 & 8.02 $\pm$ 0.37 & 7.41 $\pm$ 0.31 & 20.51 $\pm$ 1.56 \\
  & $\rho_{12}$ & \textit{U}(1.5, 8.3) $n_0$ & 8.13 $\pm$ 0.57 & 8.71 $\pm$ 0.61 & 8.45 $\pm$ 0.64 & 8.38 $\pm$ 1.64 \\
  & $\rho_{23}$ & \textit{U}($\rho_{12}$, 8.3) $n_0$ & 3.53 $\pm$ 0.43 & 6.08 $\pm$ 0.45 & 3.86 $\pm$ 0.42 & 7.64 $\pm$ 1.82 \\
  & $\log(\varepsilon_c)$ & \textit{U}(14.6, max($\varepsilon_c$)) & 982.44 $\pm$ 4.20 & 1074.48 $\pm$ 4.60 & 1089.29 $\pm$ 4.20 & 1213.18 $\pm$ 18.97 \\ \midrule
  \multicolumn{3}{l}{Log-evidence $\widehat{\ln\mathcal{Z}}$} & -1.43 $\pm$ 0.02 & -1.96 $\pm$ 0.02 & -0.92 $\pm$ 0.02 & -1.70 $\pm$ 0.02 \\
 \bottomrule
 \midrule
 CS model & K &  \textit{U}(1.7, 2.76) & 7.37 $\pm$ 0.67& 16.40 $\pm$ 0.92 & 11.32 $\pm$ 0.71 & 43.83 $\pm$ 1.16\\
 & $a_1$ & \textit{U}(0.1, 1.5) & 7.74 $\pm$ 0.64 & 6.80 $\pm$ 0.52 & 6.92 $\pm$ 0.51 & 7.63 $\pm$ 0.54 \\
 & $a_2$ & \textit{U}(1.5, 12.0) & 5.94 $\pm$ 0.64& 11.5 $\pm$ 0.88 & 10.18 $\pm$ 0.69 & 46.53 $\pm$ 1.23 \\
 & $a_3$/$a_2$ & \textit{U}(0.05, 2.0) & 42.48 $\pm$ 1.63 & 40.77 $\pm$ 1.28 & 43.54 $\pm$ 1.30 & 52.06 $\pm$ 1.46 \\
 & $a_4$ & \textit{U}(1.5, 37.0) & 20.10 $\pm$ 1.13 & 16.48 $\pm$ 0.92 & 17.43 $\pm$ 0.98 & 23.35 $\pm$ 1.00\\
 & $a_5$ & \textit{U}(0.1, 1.0) & 2.02 $\pm$ 0.31 & 2.16 $\pm$ 0.35 & 1.12 $\pm$ 0.23& 1.09 $\pm$ 0.16 \\
 & $\log(\varepsilon_c)$ & \textit{U}(14.6, max($\varepsilon_c$)) & 1331.8 $\pm$ 3.77 & 1376.7 $\pm$ 3.34 & 1440.8 $\pm$ 3.5& 1508.7 $\pm$ 3.12 \\ \midrule
  \multicolumn{3}{l}{Log-evidence $\widehat{\ln\mathcal{Z}}$} & -1.83 $\pm$ 0.02 & -2.65 $\pm$ 0.02 & -1.11 $\pm$ 0.02 & -2.46 $\pm$ 0.02 \\
 \bottomrule
 \midrule
 \multicolumn{3}{l}{Bayes' factor} & 1.49 & 1.99 & 1.21 & 2.14 \\ \bottomrule
\end{tabular}
\caption{The parameters in the PP and CS model with their corresponding priors, where $U(a, b)$ denotes uniformly sampled between $a$ and $b$. Note that the upper prior bound on the central energy densities is a variable that depends on the other EOS parameters, see Section \ref{subsec:priors}. Also shown are the estimated log-evidences $\widehat{\ln\mathcal{Z}}$ for the four models considered in this paper, and parameter-by-parameter Kullback-Leibler (KL) divergences $\widehat{D}_{\rm KL}$ (mean values and standard deviation), for each individual parameter. We report the parameter information gain for: (i) likelihood functions $L(M,R)$ associated with the Bogdanov~et~al.~(2019) $T=1.94$~Ms \NICER data set; and (ii) crudely simulated likelihood functions $L^{\dagger}(M,R)$ for an exposure time of $2T$. Errors are estimated by calculating these quantities for a set of equally-weighted realizations of the nested-sampling process, where each realization has an associated posterior distribution. The low absolute value of the KL divergence in \textit{bits} is indicative that not much information is gained over the prior in each model considered here. The Bayes' factors show that for the \texttt{ST+CST} model there is substantially more support for the PP model, caused by the stricter prior constraints on higher radii for the CS model. One should be careful in comparing evidences in combinations other than the reported Bayes' factors: the nuisance-marginalized mass-radius likelihood functions need to be normalized appropriately (by another evidence), and therefore only if a likelihood function is shared between models does this normalization cancel exactly. Evidence ratios between models with different surface hot-regions depend on evidences estimated by \citet{Riley19}; the error intervals for the \TT{ST+PST} and \TT{ST+CST} models overlap substantially, however, and can be safely equated. The normalization for a likelihood function $L^{\dagger}(M,R)$ defined by simulating extension of observing time is unknown.}
\label{table:tab1}
\end{table*}

The dense matter information yield, conditional on \NICER observations of PSR~J0030$+$0451, is weak in the context of existing knowledge, both theoretical and observational. An important aspect of any telescope mission is resource management: we therefore now investigate a scenario wherein the integrated observing time is increased. The analysis presented by \citealt{Riley19} used a data set with an integrated exposure time of $1.94$~Ms, curated by \citet{Bogdanov19a}.

Previous studies that have examined how posterior estimation of mass and radius is sensitive to factors including the number of source counts in the event data indicate that constraining power increases as the square root of the number of counts \citep{Lo13,Psaltis14b} and thus observing time $T$. It is unclear, however, whether the credible region areas should be expected to scale as $\sim\!1/\sqrt{T}$ or $\sim\!1/T$. The mass and radius are correlated in \citet[][]{Lo13} and \citet{Miller15}, but the areal reduction is given for approximate uncertainty in $M$ and $R$ separately. Let us simply suppose that the credible regions contract along some dimension $M/R\approx\rm{constant}$ by a factor of $\sim\!1/\sqrt{T}$, and also enjoy a $\sim\!1/\sqrt{T}$ scaling along the local (orthogonal) compactness direction; the overall scaling of area is then $\sim1/T$. Considering this scaling as optimistic, and a scaling of $\sim1/\sqrt{T}$ as conservative, we would require $\sim\!2$--$4$ times the exposure on PSR~J0030$+$0451 in order to halve the credible region area.

Let us make the assumption that the credible regions for mass and radius \textit{halve} in area. In Figure~\ref{fig:fig5} we show these speculative posterior distributions conditional on the \texttt{ST+PST} and \texttt{ST+CST} models, corresponding to some unknown extension to the observing time. We contract the credible regions, simply assuming that the posterior distributional mean vector is insensitive to continued observation (refer to Appendix~\ref{app:likelihood modification}). The effect is to artificially increase the absolute curvature of the mass-radius nuisance-marginalized likelihood function.

For the \texttt{ST+PST} model we note that the distributions remain very similar to the distributions in the upper panels of Figure~\ref{fig:fig3}: the $68\%$ highest-density credible region of the updated mass-radius posterior\footnote{Normalised, nuisance-marginalized likelihood function.} spans almost the entire region containing $95\%$ of the prior mass.
For the \texttt{ST+CST} model the distributions do show slightly more support for higher radii, while the range of the $68\%$ and $95\%$ credible intervals decreases. Examination of the one-dimensional KL divergences for individual parameters exhibit, in most cases, a small increase. In some cases, however, the KL divergence for individual parameters decreases, but only when the KL divergence is close to zero where the error intervals due to sampling noise substantially overlap. 
The Bayes' factors indicate that there is still only substantial posterior support for the PP model when considering the \texttt{ST+CST} model.

\section{Discussion and Conclusions}
\label{discussion}

\subsection{Prospects for PSR~J0030$+$0451 as an EOS probe}\label{sec:prospects}
Following the reported mass-radius posterior distribution from data obtained with \NICER on PSR~J0030$+$0451 by \citep{Riley19} we have explored the implications for the dense matter EOS. Two distinct hot-region models were considered, the superior \texttt{ST+PST} and also (for illustrative purposes) \texttt{ST+CST}, both yielding a different constraint on the NS radius. We have inferred the EOS using two high-density parameterizations, the PP model and CS model, which are matched to the cEFT band just above nuclear saturation density. The posterior distributions and corresponding KL divergences have shown that not much information is gained over the, already narrow, prior for either hot-region model. This can be attributed to the relatively large uncertainty in both mass-radius likelihood functions compared to the highest-density region of our prior, and the substantial overlap between the two. From the distributions shown in Figure \ref{fig:fig2} we observe that the changes from the prior to the posterior are also insignificant at nuclear densities, so that from the present analysis we cannot draw conclusions about further constraints on dense matter interactions within cEFT. 

The Bayes' factors indicate that for this particular source neither parameterization is preferred, although when using the \texttt{ST+CST} model there is slightly more support for the PP model, a result of the CS model not being able to produce large enough radii to be somewhat consistent with the likelihood function. When more information on NS radii will become available in the future we would hopefully be able to discard certain parameterizations or construct new functional EOS forms based on the information encoded in the Bayes' factors. 

In all these cases the posterior distributions are mostly informed by our prior choices, the two most stringent constraints being: (i) the lower limit of each EOS supporting a $1.97$~M$_{\odot}$ NS, which is a simple way to include the information encoded in the pulsar radio-timing (and companion modeling) mass measurement by \citet[][]{Antoniadis13}; and (ii) the computed cEFT band of possible EOS around nuclear saturation density. The first constraint causes a lower limit on the NS radius, while the latter has the opposite effect, resulting in a strongly peaked prior between $\sim\!11$--$13$~km. With the tentative measurement of a $2.17_{-0.10}^{+0.11}$~M$_{\odot}$ pulsar \citep[PSR J0740+6620;][]{Cromartie19} the prior might even be updated to further exclude EOS with small radii. We discuss the implementation of these priors in detail in Section~\ref{subsec:discuss prior implementation}.

As discussed in \citet{Riley19} there are good prospects for advancing our understanding of the \NICER background (from sources other than the MSP), and therefore tightening the PSR~J0030$+$0451 mass-radius constraint via re-analysis. Any such re-analysis may also involve more sophisticated modeling of the MSP (and its near vicinity). It is not possible---at least at present---to robustly forecast how the mass-radius nuisance-marginalized function would change in response to such modeling efforts.

Substantial extension of the PSR~J0030$+$0451 observing time is feasible given that the \NICER mission has recently been extended for three more years. In Section~\ref{sec:simulated obs time ext} we crudely simulated the evolution of the nuisance-marginalized likelihood function with observing time. Several remarks must clearly be made in regards to this: (i) we neglected the notion of re-analysis of the \NICER data curated by \citet{Bogdanov19a}; and (ii) the studies by \citet{Lo13} and \citet{Psaltis14b} on credible interval scaling assumed a single circular or infinitesimal single-temperature hot spot, not a more complex hot-region geometry such as that which emerged in \citet{Riley19}. While constraints should certainly improve with increased exposure time, we cannot confirm the precise scaling for these more complex hot-region geometries without further study, and re-analysis may yield more constraining power.

Our modification of the nuisance-marginalized likelihood function---which may require substantial future observing time to formally realize---does not promise to enhance the dense matter information yield in the context of the radio-timing pulsar mass likelihood function and of the calculations of nuclear interactions based on cEFT around the nuclear saturation density. Ultimately, in order to improve synergy with the radio-timing probe of dense matter, an independent and tight mass constraint for PSR~J0030$+$0451 would need to be combined with our compactness constraint conditional on \NICER observations. However, the pulsar is not in a binary and thus there is no known prospect of an independent tight mass constraint---based on radio observations or otherwise.

In conclusion, we cannot yet robustly justify further allocation of observing time to PSR~J0030$+$0451 for the purpose of dense matter study---at least without first exhausting modeling avenues for the $1.94$~Ms \citet{Bogdanov19a} data set, and without similar modeling of the other primary \NICER targets, principally PSR~J0437$-$4715, which has a tight mass constraint derived via radio-timing. Such a statement would arguably be valid even if the projected information gain was deemed substantial because it would be based on a crude forecast of the response to extended observing time, and the current state of knowledge may yet evolve via remodeling.

\subsection{Prior robustness and implementation approximations}\label{subsec:discuss prior implementation}

Chiral effective field theory allows for a systematic expansion of nuclear forces between neutrons and protons at low energies in terms of long-range pion-exchange contributions and short-range interactions \citet{Epelbaum:2008ga,Machleidt:2011zz}. Within cEFT it is possible to determine contributions to nucleon-nucleon and many-body forces at different orders in the low-energy expansion and to provide estimates of theoretical uncertainties due to neglected higher-order terms. While theoretical predictions for systems with a significant proton fraction, such as atomic nuclei and isospin-symmetric matter, generally depend more sensitively on properties of the employed interactions, the scheme dependence of results for pure neutron systems or very neutron-rich systems exhibit a remarkable insensitivity to such details; see \citet{Hebeler15} for a review, and also \citet{Lynn16} and \citet{Drischler19}. In addition, the theoretical results for the nuclear symmetry energy are in good agreement with experimental constraints \citep{Hebeler13}. These findings suggest that the predictions for the EOS of neutron-rich matter up to about nuclear saturation density are robust and rather well constrained. Current efforts aim at determining more systematically the upper density limit for such calculations, which is suspected to be closely related to the breakdown scale of cEFT. These studies may allow us to extend the calculations in a reliable way to higher densities and by this reduce also the EOS range of the PP and CS extensions at high densities. We note that in the analyses in this Letter we have approximated the EOS around nuclear densities with a polytropic fit to the cEFT band calculated by \citet{Hebeler13}. The true EOS within the cEFT band is however unknown, causing additional uncertainty in the presented posteriors that remains to be quantified. 

One other assumption that merits comment is the prior requirement that each EOS support a $1.97 \, M_{\odot}$ NS (Section~\ref{subsec:bayesian}). This hard binary restriction discards the information encoded in the radio pulsar mass measurement: the information in the \textit{shape} of the likelihood function is not included; and information is lost by assigning zero (rather than the correct) posterior density to EOS that do not support a $1.97 \, M_{\odot}$ star; this is not compatible with the notion of future Bayesian updating. A more accurate approach would use the pulsar mass measurement as a nuisance-marginalized likelihood function and would not truncate until far into the tails \citep{Raaijmakers18}.\footnote{To clarify with an alternative perspective, suppose that one is simultaneously combining independent likelihood functions (e.g., from different stars) in a population-level analysis. The information encoded in a subset of the other likelihood functions might strongly weight EOS that would be outside of the prior support according to a binary mass-threshold condition. Consequently, independent likelihood information would be censored.}

We considered the conclusions of this present work to be sufficiently insensitive to the above likelihood implementation detail to warrant discussion instead of recalculation. The approach taken here has the advantage that it enabled a fast, approximate separation of posterior information gain based on PSR~J0030$+$0451 from the mass information gain based on PSR~J0348$+$0432; it also avoided an additional parameter (a central density) in the sampling problem, leading to a minor reduction in computational expense.  In future applications, however, we advocate for the default interpretation of (radio) pulsar mass measurements as nuisance-marginalized likelihood functions.

\subsection{Effect of rotation on the accuracy of likelihood evaluation}
\label{subsec:rotation}

The effect of rotation on a NS is to deform it into an oblate spheroid with a larger equatorial radius and to increase the mass compared to a non-rotating NS.  These effects are correctly included in the mass and radius inference reported by \citet{Riley19}. The joint mass-radius inference made use of an empirical quasi-universal relation for the oblate shape of the surface; the surface was then embedded in the Schwarzschild metric, and higher-order metric and shape corrections were neglected as is acceptable for stars spinning near 200~Hz \citep{AlGendy2014}. In particular, the inferred radius in the work of \citet{Riley19} is the rotating star's equatorial radius. However, in this Letter, we assume that the NS is spherical, with a radius equal to the equatorial radius of the rotating star. It is worthwhile to consider the effect of this simplifying assumption on the accuracy of these results by computing sequences of rotating axisymmetric stars in hydrostatic equilibrium using the \project{RNS} code \citep{Stergioulas}.

\begin{figure*}
\centering
\includegraphics[width=\textwidth]{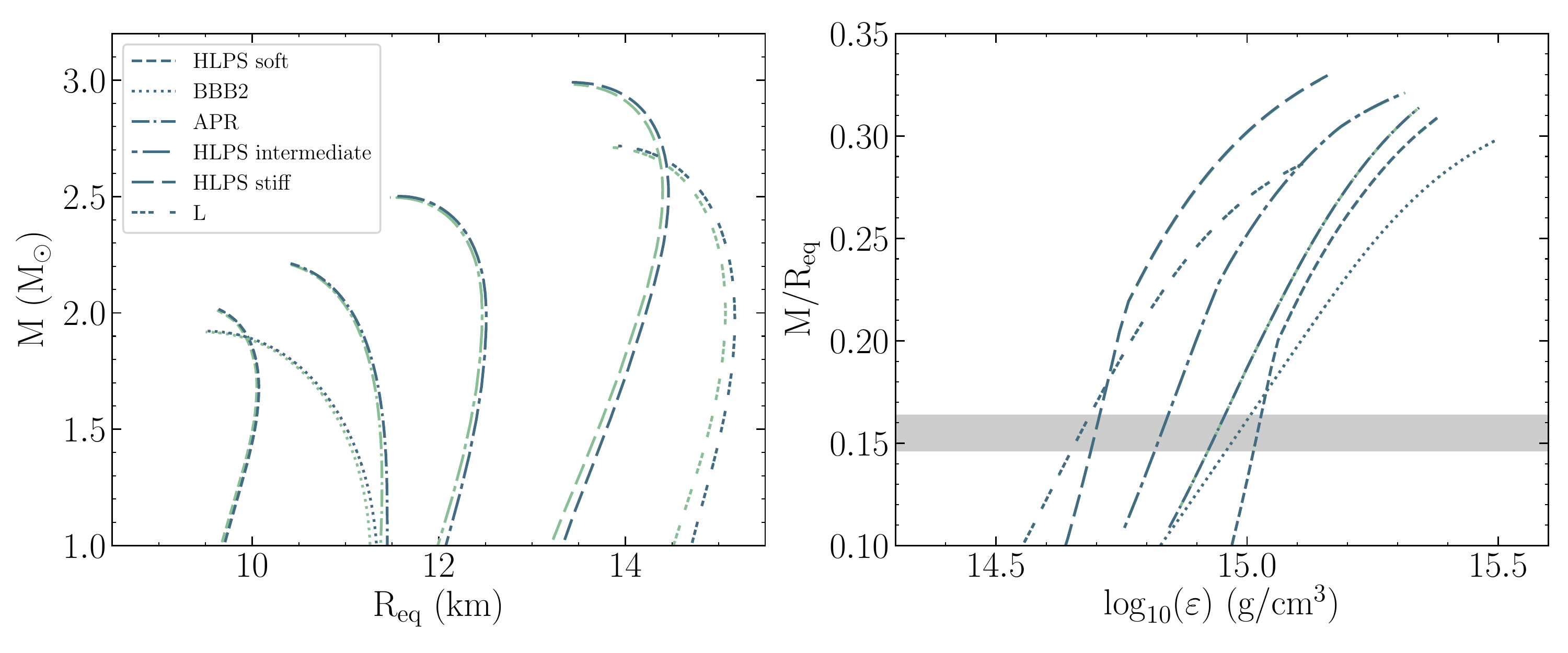}
\caption{Left panel: effect of spin on the mass versus equatorial radius curves for six representative EOS allowing for a wide range of stiffness. 
Two curves with dark and light lines are shown for each EOS. For each pair, the lighter curve with smaller radii corresponds to the zero-spin mass-radius curve, while the darker curve with larger radii is the mass-radius curve for stars spinning at 205~Hz. The EOS in order of increasing stiffness (i.e., in order of increasing radius for a $1.4$\,M$_\odot$ star) are HLPS Soft \citep{Hebeler13}, one of the softest EOS allowed by nuclear physics; BBB2 \citep{Baldo97} is a soft EOS just marginally ruled out by the observation of a 1.97\,M$_\odot$ 
pulsar; APR \citep{Akmal98} includes boost corrections; HLPS Intermediate and Stiff are representative EOS from \citet{Hebeler13}; and L
\citep{Pandharipande75}, a very stiff EOS, is most likely ruled out by the Laser Interferometer Gravitational-Wave Observatory (LIGO) observation of GW170817 \citep{GW170817discovery}. The HLPS EOS are also shown in Figures \ref{fig:fig1} and \ref{fig:fig2}.
Right panel: effect of spin on the equatorial compactness ratio $M/R$ versus central energy density. 
The order of curves from left to right at a value of $M/R_{\rm eq}=0.15$ is from stiffest to softest EOS. Two curves, corresponding to 0 and 205~Hz are plotted for each EOS, however the difference between the curves is smaller than the line width so it is difficult to see the difference by eye. The gray horizontal box shows the compactness range of $M/R_{\rm eq} = 0.156^{+0.008}_{-0.010}$ for the \TT{ST+PST} model reported in \citet{Riley19}.}
\label{fig:mrspin}
\end{figure*}

The pulsar PSR~J0030$+$0451 rotates with a spin frequency of 205~Hz. Figure \ref{fig:mrspin} demonstrates the effect of rotation at this spin frequency on some representative EOS. In the left panel of Figure~\ref{fig:mrspin} we plot mass versus equatorial radius curves for stars spinning at 0 and 205~Hz for a representative set of tabulated EOS that cover a wide range of stiffness (see figure caption). This shows that the increase in radius at fixed mass is at most 0.2 km for the stiffest EOS.  More physically plausible EOS, such as the representative EOS of \citet{Hebeler13}, are deformed by a smaller coordinate distance. The difference of 0.2 km is much smaller than the precision, $\pm 1.2$ km, of the \textit{marginal} radius estimate in \citet{Riley19}. The EOS constructed in this Letter using the PP and CS models will respond to rotation in a similar fashion. This suggests that this analysis is insensitive to the effects of rotation on the properties of this pulsar.

If two stars with the same central energy density and EOS but different spin rates are compared, the mass of the faster rotating star is larger \citep[e.g.,][]{Hartle1967,HT1968}. Given that the tightest constraints reported by \citep{Riley19} are on the equatorial compactness ratio of $M/R$, it is more constructive to compute the changes in the equatorial compactness ratio, instead of the mass or radius, as a star spins. For zero-spin stars, given an EOS, each possible value of central energy density is mapped by the equations of hydrostatic equilibrium to a unique compactness ratio. The right panel of Figure \ref{fig:mrspin} shows the curves of compactness versus central energy density for the same representative set of six different EOS.\footnote{Note that for the three representative EOS of \citet{Hebeler13} there are some values of central energy density where the slope of the compactness curve changes. These values of density correspond where the piecewise polytropes are matched in these EOS.}
Note that Figure \ref{fig:mrspin} actually shows 12 curves, corresponding to two different spin rates of 0 and 205~Hz for each EOS. The curves for 205~Hz differ from the zero-spin curves by an amount that is smaller than 0.1\%, which is smaller than the thickness of the line. 
This suggests that if the compactness for a particular central energy density and EOS is computed, it does not matter whether or not spin is included in the calculation. This property appears to extend to higher values of spin and will be investigated in more detail elsewhere.

\subsection{Consistency with previous EOS constraints}
There have been several attempts to constrain the parameters of dense matter EOS models using joint posterior information about mass and radius, where that information was derived via phase-averaged X-ray spectral modeling of bursting and quiescent accreting NSs (see Section 4 of \citealt{Riley19}). \citet{Bogdanov16} and \citet{Ozel16} used a piecewise-polytropic EOS model, inferring both EOS parameters and the associated mass-radius bands. \citet{Steiner10,Steiner13,Steiner18} and \citet{Lattimer14} considered a larger range of parameterized EOS models to infer dense matter parameters including those associated with the nuclear symmetry energy and the mass-radius bands. \citet{Baillot19} have performed similar analysis using the parameterized EOS model of \citet{Margueron18a,Margueron18b}, inferring symmetry energy parameters, speed of sound profiles, and mass-radius bands.

A direct comparison between the EOS constraints derived in these works and ours is difficult: they use different models and priors, and the influence of the priors is not always clear from the published analysis. Due to the form of the models being used, we expect that the models used in these publications should also have a clear peak in the prior distributions in both pressure-energy density and mass-radius space.  An exception is Model C of \citet{Steiner13}, which is formulated to give a flat prior distribution in pressure-energy density space (although it may still have a peak in the mass-radius space prior distribution due to the non-linear nature of the Tolman-Oppenheimer-Volkoff  equation mapping).  That the EOS model and priors have a major influence is clear from Figure 4 of \citet{Steiner13}, which shows the inferred posterior distributions for one of the symmetry energy parameters.  There is clear variation between models, and the posterior for Model C is noticeably broader. To perform a consistent comparison with the EOS parameters inferred in our work, we would need to know the prior distributions on the pressure-energy density and mass-radius space associated with the EOS parameterizations and priors used in these previous works (the equivalent, for those models, of our Figure \ref{fig:fig1}).  However, the necessary details are not shown in these earlier works, so further study will be required to make a robust comparison.

Recently, the gravitational wave (GW) observation of the NS binary inspiral GW170817 \citep{GW170817discovery}, where the progenitor is widely accepted to be a NS-NS system, has provided an independent method for constraining the EOS by measuring tidal effects of the NS in the evolution of GW phase. The dominant tidal GW imprint depends on the characteristic tidal deformability parameters $\Lambda=(2/3)k_2 (c^2 R_{\rm eq})^5/(G M)^{5}$, where $k_2$ is the EOS-dependent Love number \citep{Flanagan:2007ix}.
Several studies, as discussed below, infer $\Lambda$ from the GW data (e.g., \citealt{GW170817discovery,De18,GW170817sourceproperties, Abbott18}). For example, work performed by the LIGO-Virgo Scientific Collaborations \citep{GW170817sourceproperties} quantifies the impact of the choice of spin priors and systematic uncertainties in the waveform models, and find these to be non-negligible yet smaller than the statistical errors. Specializing to a low-spin prior with a dimensionless value of less than 0.05 (as expected from extrapolating the spin-down of observed Galactic binary pulsars that will merge within a Hubble time), a representative GW model, and the case where both binary components are assumed to be NSs and have the same EOS, \cite{Abbott18} inferred constraints on the EOS and the radius using two methods: a parameterized EOS and approximate EOS-insensitive relations. Although the inferred masses from the GW data involved in GW170817 \citep{GW170817discovery,GW170817sourceproperties} were similar to PSR~J0030$+$0451 within the measurement uncertainties, the compatibility of EOS results is difficult to assess in detail because of the different priors imposed in the analysis. However, the inferred $90\%$ credible intervals for the radii of the two components $R\in [9.1, 12.8]$\,km and $R \in [10.5 ,13.3]$\,km with the two methods are consistent with the results in this Letter.
Independent analyses of GW170817-only data \citep{De18} and results (e.g., \citealt{Annala18,Lim:2018bkq,Most:2018hfd,Raithel18,Tews:2018chv}) obtained compatible constraints of the radius in a broad range of $R\in [9, 14]$\,km. 
With the ongoing third observing run of the Laser Interferometer Gravitational-Wave Observatory (LIGO) and Virgo detectors at a higher detector sensitivity and further improvements planned \citep{Aasi:2013wya, LIGOScientific:2019vkc}, GW measurements of a greater number of NS binaries (encompassing both NS-NS and NS-black hole systems) will yield tighter EOS constraints in the coming years. 

\subsection{Final remarks}
We have studied the implications of the available PSR~J0030$+$0451 mass-radius likelihood information for dense matter EOS knowledge. The likelihood function of mass and radius is predominantly sensitive to their combination in the compactness ratio, conditional on the current \NICER data set and X-ray pulse-profile modeling.  The posterior information gain over our choice of prior knowledge is weak in the joint context of both prior constraints imposed by cEFT interactions at nuclear densities, and all EOS being able to support a 1.97\,M$_\odot$ NS. This is a consequence of the substantial overlap between the relatively broad mass-radius likelihood function and our narrowly peaked prior.
However, we have shown how our methods can be applied to data obtained through pulse-profile modeling of MSPs. Our understanding of the nature of dense matter is expected to improve in the near future with the constraining power offered by the \NICER mission: \NICER is concurrently observing rotation-powered MSPs such as PSR~J0437$-$4715 that have an independent mass measurement derived via radio timing.
Moreover, joint radio and X-ray information from these MSPs promises synergism with the radio information from high-mass pulsars such as PSR~J0348$+$0432. 

\acknowledgments
This work was supported in part by NASA through the \NICER mission and the Astrophysics Explorers Program. T.E.R. and A.L.W. acknowledge support from ERC Starting Grant No.~639217 CSINEUTRONSTAR (PI: Watts). A.L.W. would like to thank Andrew Steiner for useful discussions on the role of priors in previously published results. The authors would also like to thank Kent Wood for helpful comments. This work was sponsored by NWO Exact and Natural Sciences for the use of supercomputer facilities, and was carried out on the Dutch national e-infrastructure with the support of SURF Cooperative. S.K.G., K.H., and A.S. acknowledge support from the DFG through SFB 1245. G.R., T.H., and S.N. are grateful for support from the Nederlandse Organisatie voor Wetenschappelijk Onderzoek (NWO) through the VIDI and Projectruimte grants (PI: Nissanke). S.G. acknowledges the support of the CNES. S.M.M. thanks NSERC for support. J.M.L. acknowledges support from NASA through Grant 80NSSC17K0554 and the U.S. DOE from Grant DE-FG02-87ER40317. R.M.L. acknowledges the support of NASA through Hubble Fellowship Program grant HST-HF2-51440.001. This research has made extensive use of NASA's Astrophysics Data System Bibliographic Services (ADS) and the arXiv. 

\software{Python/C~language~\citep{python2007}, GNU~Scientific~Library~\citep[GSL;][]{Gough:2009}, NumPy~\citep{Numpy2011}, Cython~\citep{cython2011}, SciPy~\citep{Scipy}, MPI~\citep{MPI}, \project{MPI for Python}~\citep{mpi4py}, Matplotlib~\citep{Hunter:2007,matplotlibv2}, IPython~\citep{IPython2007}, Jupyter~\citep{Kluyver:2016aa}, \MultiNest~\citep{Feroz09}, \textsc{PyMultiNest}~\citep{Buchner14}, 
\project{RNS}~\citep[][]{Stergioulas}.}

\bibliographystyle{aasjournal}
\bibliography{references.bib}

\begin{thebibliography}{}
\expandafter\ifx\csname natexlab\endcsname\relax\def\natexlab#1{#1}\fi
\providecommand{\url}[1]{\href{#1}{#1}}

\bibitem[{{Abbott} {et~al.}(2017){Abbott}, {Abbott}, {Abbott}, {Acernese},
  {Ackley}, {Adams}, {Adams}, {Addesso}, {Adhikari}, {Adya}, \&
  et~al.}]{GW170817discovery}
{Abbott}, B.~P., {Abbott}, R., {Abbott}, T.~D., {et~al.} 2017, Physical Review
  Letters, 119, 161101

\bibitem[{{Abbott} {et~al.}(2018){Abbott}, {Abbott}, {Abbott}, {Acernese},
  {Ackley}, {Adams}, {Adams}, {Addesso}, {Adhikari}, {Adya}, \&
  et~al.}]{Abbott18}
---. 2018, \prl, 121, 161101

\bibitem[{Abbott {et~al.}(2018)}]{Aasi:2013wya}
Abbott, B.~P., {et~al.} 2018, Living Rev. Rel., 21, 3

\bibitem[{{Abbott} {et~al.}(2019){Abbott}, {Abbott}, {Abbott}, {Acernese},
  {Ackley}, {Adams}, {Adams}, {Addesso}, {Adhikari}, {Adya}, \&
  et~al.}]{GW170817sourceproperties}
{Abbott}, B.~P., {Abbott}, R., {Abbott}, T.~D., {et~al.} 2019, Physical Review
  X, 9, 011001

\bibitem[{{Akmal} {et~al.}(1998){Akmal}, {Pandharipande}, \&
  {Ravenhall}}]{Akmal98}
{Akmal}, A., {Pandharipande}, V.~R., \& {Ravenhall}, D.~G. 1998, \prc, 58, 1804

\bibitem[{{Alford} \& {Han}(2016)}]{Alford16}
{Alford}, M.~G., \& {Han}, S. 2016, European Physical Journal A, 52, 62

\bibitem[{{AlGendy} \& {Morsink}(2014)}]{AlGendy2014}
{AlGendy}, M., \& {Morsink}, S.~M. 2014, ApJ, 791, 78

\bibitem[{{Annala} {et~al.}(2018){Annala}, {Gorda}, {Kurkela}, \&
  {Vuorinen}}]{Annala18}
{Annala}, E., {Gorda}, T., {Kurkela}, A., \& {Vuorinen}, A. 2018, \prl, 120,
  172703

\bibitem[{{Antoniadis} {et~al.}(2013){Antoniadis}, {Freire}, {Wex}, {Tauris},
  {Lynch}, {van Kerkwijk}, {Kramer}, {Bassa}, {Dhillon}, {Driebe}, {Hessels},
  {Kaspi}, {Kondratiev}, {Langer}, {Marsh}, {McLaughlin}, {Pennucci}, {Ransom},
  {Stairs}, {van Leeuwen}, {Verbiest}, \& {Whelan}}]{Antoniadis13}
{Antoniadis}, J., {Freire}, P. C.~C., {Wex}, N., {et~al.} 2013, Science, 340,
  448

\bibitem[{{Baillot d'Etivaux} {et~al.}(2019){Baillot d'Etivaux}, {Guillot},
  {Margueron}, {Webb}, {Catelan}, \& {Reisenegger}}]{Baillot19}
{Baillot d'Etivaux}, N., {Guillot}, S., {Margueron}, J., {et~al.} 2019, arXiv
  e-prints, arXiv:1905.01081

\bibitem[{{Baldo} {et~al.}(1997){Baldo}, {Bombaci}, \& {Burgio}}]{Baldo97}
{Baldo}, M., {Bombaci}, I., \& {Burgio}, G.~F. 1997, \aap, 328, 274

\bibitem[{{Baym} {et~al.}(2018){Baym}, {Hatsuda}, {Kojo}, {Powell}, {Song}, \&
  {Takatsuka}}]{Baym18}
{Baym}, G., {Hatsuda}, T., {Kojo}, T., {et~al.} 2018, Reports on Progress in
  Physics, 81, 056902

\bibitem[{Baym {et~al.}(1971)Baym, Pethick, \& Sutherland}]{Baym71}
Baym, G., Pethick, C., \& Sutherland, P. 1971, Astrophys. J., 170, 299

\bibitem[{{Behnel} {et~al.}(2011){Behnel}, {Bradshaw}, {Citro}, {Dalcin},
  {Seljebotn}, \& {Smith}}]{cython2011}
{Behnel}, S., {Bradshaw}, R., {Citro}, C., {et~al.} 2011, Computing in Science
  Engineering, 13, 31
  
  \bibitem[{{Bilous} {et~al.}(2019){Bilous}, {Watts}, {Harding}, {Riley}, {Arzoumanian}, {Bogdanov}, {Gendreau}, {Ray}, {Guillot}, {Ho}, \&
  {Chakrabarty}}]{Bilous19}
{Bilous}, A.~V., {Watts}, A.~L., {Harding}, A.K., {et~al.}, 2019,
  \apjl, 887, L23

\bibitem[{{Bogdanov}(2016)}]{Bogdanov16b}
{Bogdanov}, S. 2016, European Physical Journal A, 52, 37

\bibitem[{{Bogdanov} {et~al.}(2016){Bogdanov}, {Heinke}, {{\"O}zel}, \&
  {G{\"u}ver}}]{Bogdanov16}
{Bogdanov}, S., {Heinke}, C.~O., {{\"O}zel}, F., \& {G{\"u}ver}, T. 2016, \apj,
  831, 184

\bibitem[{{Bogdanov} {et~al.}(2019){Bogdanov}, {Guillot}, {Ray}, {Wolff},
  {Chakrabarty}, {Ho}, {Kerr}, {Lamb}, {Lommen}, {Ludlam}, {Milburn},
  {Montano}, {Miller}, {Baub\"ock}, {\"Ozel}, {Psaltis}, {Remillard}, {Riley},
  {Steiner}, {Strohmayer}, {Watts}, {Wood}, {Zeldes}, {Enoto}, {Okajima},
  {Kellogg}, {Baker}, {Markwardt}, {Arzoumanian}, \& {Gendreau}}]{Bogdanov19a}
{Bogdanov}, S., {Guillot}, S., {Ray}, P.~S., {et~al.}, \apjl, 887, L25

\bibitem[{{Buchner} {et~al.}(2014){Buchner}, {Georgakakis}, {Nandra}, {Hsu},
  {Rangel}, {Brightman}, {Merloni}, {Salvato}, {Donley}, \&
  {Kocevski}}]{Buchner14}
{Buchner}, J., {Georgakakis}, A., {Nandra}, K., {et~al.} 2014, \aap, 564, A125

\bibitem[{{Cromartie} {et~al.}(2019){Cromartie}, {Fonseca}, {Ransom},
  {Demorest}, {Arzoumanian}, {Blumer}, {Brook}, {DeCesar}, {Dolch}, {Ellis},
  {Ferdman}, {Ferrara}, {Garver-Daniels}, {Gentile}, {Jones}, {Lam}, {Lorimer},
  {Lynch}, {McLaughlin}, {Ng}, {Nice}, {Pennucci}, {Spiewak}, {Stairs},
  {Stovall}, {Swiggum}, \& {Zhu}}]{Cromartie19}
{Cromartie}, H.~T., {Fonseca}, E., {Ransom}, S.~M., {et~al.} 2019, Nature
  Astronomy, 439

\bibitem[{Dalc\'{i}n {et~al.}(2008)Dalc\'{i}n, Paz, Storti, \&
  D'El\'{i}a}]{mpi4py}
Dalc\'{i}n, L., Paz, R., Storti, M., \& D'El\'{i}a, J. 2008, Journal of
  Parallel and Distributed Computing, 68, 655.
\newblock
  \url{http://www.sciencedirect.com/science/article/pii/S0743731507001712}

\bibitem[{{De} {et~al.}(2018){De}, {Finstad}, {Lattimer}, {Brown}, {Berger}, \&
  {Biwer}}]{De18}
{De}, S., {Finstad}, D., {Lattimer}, J.~M., {et~al.} 2018, \prl, 121, 091102

\bibitem[{Drischler {et~al.}(2019)Drischler, Hebeler, \& Schwenk}]{Drischler19}
Drischler, C., Hebeler, K., \& Schwenk, A. 2019, Phys. Rev. Lett., 122, 042501

\bibitem[{Droettboom {et~al.}(2018)Droettboom, Caswell, Hunter, Firing,
  Nielsen, Lee, de~Andrade, Varoquaux, Stansby, Root, Elson, Dale, Lee, May,
  Seppänen, Klymak, McDougall, Straw, Hobson, cgohlke, Yu, Ma, Vincent,
  Silvester, Moad, Katins, Kniazev, Hoffmann, Ariza, \& Würtz}]{matplotlibv2}
Droettboom, M., Caswell, T.~A., Hunter, J., {et~al.} 2018,
  matplotlib/matplotlib v2.2.2, , , doi:10.5281/zenodo.1202077.
\newblock \url{https://doi.org/10.5281/zenodo.1202077}

\bibitem[{Epelbaum {et~al.}(2009)Epelbaum, Hammer, \&
  Meissner}]{Epelbaum:2008ga}
Epelbaum, E., Hammer, H.-W., \& Meissner, U.-G. 2009, Rev. Mod. Phys., 81, 1773

\bibitem[{{Feroz} \& {Hobson}(2008)}]{Feroz08}
{Feroz}, F., \& {Hobson}, M.~P. 2008, \mnras, 384, 449

\bibitem[{{Feroz} {et~al.}(2009){Feroz}, {Hobson}, \& {Bridges}}]{Feroz09}
{Feroz}, F., {Hobson}, M.~P., \& {Bridges}, M. 2009, \mnras, 398, 1601

\bibitem[{{Feroz} {et~al.}(2013){Feroz}, {Hobson}, {Cameron}, \&
  {Pettitt}}]{Feroz13}
{Feroz}, F., {Hobson}, M.~P., {Cameron}, E., \& {Pettitt}, A.~N. 2013, ArXiv
  e-prints, arXiv:1306.2144

\bibitem[{Flanagan \& Hinderer(2008)}]{Flanagan:2007ix}
Flanagan, E.~E., \& Hinderer, T. 2008, Phys. Rev., D77, 021502

\bibitem[{Forum(1994)}]{MPI}
Forum, M.~P. 1994, MPI: A Message-Passing Interface Standard, Tech. rep.,
  Knoxville, TN, USA

\bibitem[{Fraga {et~al.}(2014)Fraga, Kurkela, \& Vuorinen}]{Fraga14}
Fraga, E.~S., Kurkela, A., \& Vuorinen, A. 2014, Astrophys. J., 781, L25

\bibitem[{{Gendreau} {et~al.}(2016){Gendreau}, {Adkins}, {et~al.}}]{Gendreau16}
{Gendreau}, K.~C.~{Arzoumanian}, Z., {Adkins}, P.~W., {et~al.} 2016, in
  Proceedings of SPIE, Vol. 9905, Space Telescopes and Instrumentation 2016:
  Ultraviolet to Gamma Ray, 99051H

\bibitem[{Gough(2009)}]{Gough:2009}
Gough, B. 2009, GNU Scientific Library Reference Manual - Third Edition, 3rd
  edn. (Network Theory Ltd.)

\bibitem[{{Greif} {et~al.}(2019){Greif}, {Raaijmakers}, {Hebeler}, {Schwenk},
  \& {Watts}}]{Greif19}
{Greif}, S.~K., {Raaijmakers}, G., {Hebeler}, K., {Schwenk}, A., \& {Watts},
  A.~L. 2019, MNRAS, 485, 5363

\bibitem[{{Harding} \& {Muslimov}(2002)}]{Harding02}
{Harding}, A.~K., \& {Muslimov}, A.~G. 2002, \apj, 568, 862

\bibitem[{{Hartle}(1967)}]{Hartle1967}
{Hartle}, J.~B. 1967, \apj, 150, 1005

\bibitem[{{Hartle} \& {Thorne}(1968)}]{HT1968}
{Hartle}, J.~B., \& {Thorne}, K.~S. 1968, ApJ, 153, 807

\bibitem[{{Hebeler} {et~al.}(2015){Hebeler}, {Holt}, {Men{\'e}ndez}, \&
  {Schwenk}}]{Hebeler15}
{Hebeler}, K., {Holt}, J.~D., {Men{\'e}ndez}, J., \& {Schwenk}, A. 2015, Annu.
  Rev. Nucl. Part. Sci., 65, 457

\bibitem[{Hebeler {et~al.}(2013)Hebeler, Lattimer, Pethick, \&
  Schwenk}]{Hebeler13}
Hebeler, K., Lattimer, J.~M., Pethick, C.~J., \& Schwenk, A. 2013, The
  Astrophysical Journal, 773, 11.
\newblock \url{https://doi.org/10.1088%2F0004-637x%2F773%2F1%2F11}

\bibitem[{Hebeler \& Schwenk(2010)}]{Hebeler10a}
Hebeler, K., \& Schwenk, A. 2010, Phys. Rev. C, 82, 014314.
\newblock \url{https://link.aps.org/doi/10.1103/PhysRevC.82.014314}

\bibitem[{{Higson}(2018)}]{Higson_nestcheck}
{Higson}, E. 2018, The Journal of Open Source Software, 3, 916

\bibitem[{Higson {et~al.}(2018)Higson, Handley, Hobson, \&
  Lasenby}]{higson2018sampling}
Higson, E., Handley, W., Hobson, M., \& Lasenby, A. 2018, Bayesian Analysis,
  13, 873.
\newblock \url{https://doi.org/10.1214/17-BA1075}

\bibitem[{Higson {et~al.}(2019)Higson, Handley, Hobson, \&
  Lasenby}]{higson2019diagnostic}
---. 2019, Monthly Notices of the Royal Astronomical Society, 483, 2044.
\newblock \url{http://doi.org/10.1093/mnras/sty3090}

\bibitem[{Hunter(2007)}]{Hunter:2007}
Hunter, J.~D. 2007, Computing in Science \& Engineering, 9, 90.
\newblock \url{http://dx.doi.org/10.1109/MCSE.2007.55}

\bibitem[{Jones {et~al.}(2001--)Jones, Oliphant, Peterson, {et~al.}}]{Scipy}
Jones, E., Oliphant, T., Peterson, P., {et~al.} 2001--, {SciPy}: Open source
  scientific tools for {Python}, , , [Online; accessed 21.06.2019].
\newblock \url{http://www.scipy.org/}

\bibitem[{Kass \& Raftery(1995)}]{KassRaft95}
Kass, R.~E., \& Raftery, A.~E. 1995, Journal of the American Statistical
  Association, 90, 773.
\newblock \url{http://www.jstor.org/stable/2291091}

\bibitem[{Kluyver {et~al.}(2016)Kluyver, Ragan-Kelley, P{\'e}rez, Granger,
  Bussonnier, Frederic, Kelley, Hamrick, Grout, Corlay, Ivanov, Avila, Abdalla,
  \& Willing}]{Kluyver:2016aa}
Kluyver, T., Ragan-Kelley, B., P{\'e}rez, F., {et~al.} 2016, in Positioning and
  Power in Academic Publishing: Players, Agents and Agendas, ed. F.~Loizides \&
  B.~Schmidt, IOS Press, 87 -- 90

\bibitem[{Kullback \& Leibler(1951)}]{kullback1951}
Kullback, S., \& Leibler, R.~A. 1951, Ann. Math. Statist., 22, 79.
\newblock \url{https://doi.org/10.1214/aoms/1177729694}

\bibitem[{{Lattimer} \& {Prakash}(2016)}]{Lattimer16}
{Lattimer}, J.~M., \& {Prakash}, M. 2016, Physics Reports, 621, 127

\bibitem[{{Lattimer} \& {Steiner}(2014)}]{Lattimer14}
{Lattimer}, J.~M., \& {Steiner}, A.~W. 2014, \apj, 784, 123

\bibitem[{Lim \& Holt(2018)}]{Lim:2018bkq}
Lim, Y., \& Holt, J.~W. 2018, Phys. Rev. Lett., 121, 062701

\bibitem[{{Lo} {et~al.}(2013){Lo}, {Miller}, {Bhattacharyya}, \& {Lamb}}]{Lo13}
{Lo}, K.~H., {Miller}, M.~C., {Bhattacharyya}, S., \& {Lamb}, F.~K. 2013, ApJ,
  776, 19

\bibitem[{Lynn {et~al.}(2016)Lynn, Tews, Carlson, Gandolfi, Gezerlis, Schmidt,
  \& Schwenk}]{Lynn16}
Lynn, J.~E., Tews, I., Carlson, J., {et~al.} 2016, Phys. Rev. Lett., 116,
  062501

\bibitem[{Machleidt \& Entem(2011)}]{Machleidt:2011zz}
Machleidt, R., \& Entem, D.~R. 2011, Phys. Rept., 503, 1

\bibitem[{{Margueron} {et~al.}(2018{\natexlab{a}}){Margueron}, {Hoffmann
  Casali}, \& {Gulminelli}}]{Margueron18a}
{Margueron}, J., {Hoffmann Casali}, R., \& {Gulminelli}, F. 2018{\natexlab{a}},
  \prc, 97, 025805

\bibitem[{{Margueron} {et~al.}(2018{\natexlab{b}}){Margueron}, {Hoffmann
  Casali}, \& {Gulminelli}}]{Margueron18b}
---. 2018{\natexlab{b}}, \prc, 97, 025806

\bibitem[{{Miller} {et~al.}(2019a){Miller}, {Chirenti}, \& {Lamb}}]{Miller2019}
{Miller}, M.~C., {Chirenti}, C., \& {Lamb}, F.~K. 2019, arXiv e-prints,
  arXiv:1904.08907
  
  \bibitem[{{Miller} {et~al.}(2019b){Miller}, {Lamb}, \&
  {Dittmann}}]{Miller19b}
{Miller}, M.~C., {Lamb}, F.K., \& {Dittmann}, A.J. {et~al.}, \apjl, 887, L24

\bibitem[{{Miller} \& {Lamb}(2015)}]{Miller15}
{Miller}, M.~C., \& {Lamb}, F.~K. 2015, Astrophysical Journal, 808, 31

\bibitem[{Most {et~al.}(2018)Most, Weih, Rezzolla, \&
  Schaffner-Bielich}]{Most:2018hfd}
Most, E.~R., Weih, L.~R., Rezzolla, L., \& Schaffner-Bielich, J. 2018, Phys.
  Rev. Lett., 120, 261103

\bibitem[{{Oertel} {et~al.}(2017){Oertel}, {Hempel}, {Kl{\"a}hn}, \&
  {Typel}}]{Oertel17}
{Oertel}, M., {Hempel}, M., {Kl{\"a}hn}, T., \& {Typel}, S. 2017, Reviews of
  Modern Physics, 89, 015007

\bibitem[{{Oliphant}(2007)}]{python2007}
{Oliphant}, T.~E. 2007, Computing in Science Engineering, 9, 10

\bibitem[{{{\"O}zel} {et~al.}(2016){{\"O}zel}, {Psaltis}, {G{\"u}ver}, {Baym},
  {Heinke}, \& {Guillot}}]{Ozel16}
{{\"O}zel}, F., {Psaltis}, D., {G{\"u}ver}, T., {et~al.} 2016, ApJ, 820, 28

\bibitem[{{Pandharipande} \& {Smith}(1975)}]{Pandharipande75}
{Pandharipande}, V.~R., \& {Smith}, R.~A. 1975, \nphysa, 237, 507

\bibitem[{{Paschalidis} \& {Stergioulas}(2017)}]{Paschalidis17}
{Paschalidis}, V., \& {Stergioulas}, N. 2017, Living Reviews in Relativity, 20,
  7

\bibitem[{{Perez} \& {Granger}(2007)}]{IPython2007}
{Perez}, F., \& {Granger}, B.~E. 2007, Computing in Science Engineering, 9, 21

\bibitem[{{Psaltis} {et~al.}(2014){Psaltis}, {{\"O}zel}, \&
  {Chakrabarty}}]{Psaltis14b}
{Psaltis}, D., {{\"O}zel}, F., \& {Chakrabarty}, D. 2014, ApJ, 787, 136

\bibitem[{{Raaijmakers} {et~al.}(2018){Raaijmakers}, {Riley}, \&
  {Watts}}]{Raaijmakers18}
{Raaijmakers}, G., {Riley}, T.~E., \& {Watts}, A.~L. 2018, MNRAS, 478, 2177

\bibitem[{{Raithel} {et~al.}(2018){Raithel}, {{\"O}zel}, \&
  {Psaltis}}]{Raithel18}
{Raithel}, C.~A., {{\"O}zel}, F., \& {Psaltis}, D. 2018, \apjl, 857, L23

\bibitem[{{Riley} {et~al.}(2018){Riley}, {Raaijmakers}, \& {Watts}}]{Riley18}
{Riley}, T.~E., {Raaijmakers}, G., \& {Watts}, A.~L. 2018, \mnras, 478, 1093

\bibitem[{{Riley} {et~al.}(2019){Riley}, {Watts}, {Bogdanov}, {Ray},
  {Ludlam}, {Guillot}, {Arzoumanian}, {Baker}, {Bilous}, {Chakrabarty},
  {Gendreau}, {Harding}, {Ho}, {Lattimer}, {Morsink}, \&
  {Strohmayer}}]{Riley19}
{Riley}, T.~E., {Watts}, A.~L., {Bogdanov}, S., {et~al.}, \apjl, 887, L21

\bibitem[{Scott(1992)}]{Scott92}
Scott, D. 1992, Multivariate Density Estimation: Theory, Practice, and
  Visualization, A Wiley-interscience publication (Wiley).
\newblock \url{https://books.google.nl/books?id=7crCUS\_F2ocC}

\bibitem[{{Shoemaker} \& {LIGO Scientific
  Collaboration}(2019)}]{LIGOScientific:2019vkc}
{Shoemaker}, D., \& {LIGO Scientific Collaboration}. 2019, \baas, 51, 452

\bibitem[{Skilling(2006)}]{skilling2006}
Skilling, J. 2006, Bayesian Anal., 1, 833.
\newblock \url{https://doi.org/10.1214/06-BA127}

\bibitem[{{Steiner} {et~al.}(2018){Steiner}, {Heinke}, {Bogdanov}, {Li}, {Ho},
  {Bahramian}, \& {Han}}]{Steiner18}
{Steiner}, A.~W., {Heinke}, C.~O., {Bogdanov}, S., {et~al.} 2018, \mnras, 476,
  421

\bibitem[{{Steiner} {et~al.}(2010){Steiner}, {Lattimer}, \&
  {Brown}}]{Steiner10}
{Steiner}, A.~W., {Lattimer}, J.~M., \& {Brown}, E.~F. 2010, ApJ, 722, 33

\bibitem[{{Steiner} {et~al.}(2013){Steiner}, {Lattimer}, \&
  {Brown}}]{Steiner13}
---. 2013, ApJ Letters, 765, L5

\bibitem[{{Stergioulas} \& {Friedman}(1995)}]{Stergioulas}
{Stergioulas}, N., \& {Friedman}, J.~L. 1995, \apj, 444, 306

\bibitem[{{Tews} {et~al.}(2018){Tews}, {Carlson}, {Gandolfi}, \&
  {Reddy}}]{Tews18b}
{Tews}, I., {Carlson}, J., {Gandolfi}, S., \& {Reddy}, S. 2018, \apj, 860, 149

\bibitem[{Tews {et~al.}(2018)Tews, Margueron, \& Reddy}]{Tews:2018chv}
Tews, I., Margueron, J., \& Reddy, S. 2018, Phys. Rev., C98, 045804

\bibitem[{{van der Walt} {et~al.}(2011){van der Walt}, {Colbert}, \&
  {Varoquaux}}]{Numpy2011}
{van der Walt}, S., {Colbert}, S.~C., \& {Varoquaux}, G. 2011, Computing in
  Science Engineering, 13, 22

\bibitem[{{Watts}(2019)}]{Watts19b}
{Watts}, A.~L. 2019, arXiv e-prints, arXiv:1904.07012

\bibitem[{{Watts} {et~al.}(2016){Watts}, {Andersson}, {Chakrabarty}, {Feroci},
  {Hebeler}, {Israel}, {Lamb}, {Miller}, {Morsink}, {{\"O}zel}, {Patruno},
  {Poutanen}, {Psaltis}, {Schwenk}, {Steiner}, {Stella}, {Tolos}, \& {van der
  Klis}}]{Watts16}
{Watts}, A.~L., {Andersson}, N., {Chakrabarty}, D., {et~al.} 2016, Reviews of
  Modern Physics, 88, 021001

\end{thebibliography}

\appendix
\section{Likelihood function modification}\label{app:likelihood modification}
Here we give the prescription for modifying the likelihood function to crudely simulate the effect of increased exposure time. Let
\begin{equation*}
L^{\dagger}(M,R)\propto L(M,R)\frac{p^{\dagger}(M,R\;|\;\bm{d})}{p(M,R\;|\;\bm{d})},
\end{equation*}
where $p^{\dagger}(M,R\;|\;\bm{d})$ is given by isotropic compression of mass with flat $p(M,R)$:
\begin{equation*}
X=\mathop{\int}_{\mathcal{C}^{\dagger}}p^{\dagger}(M,R\:|\;\bm{d})dMdR=\mathop{\int}_{\mathcal{C}}p(M,R\;|\;\bm{d})dMdR,
\end{equation*}
such that $\forall X\leq 1$, region $\mathcal{C}^{\dagger}$ has Euclidean area $A^{\dagger}$ and $A^{\dagger}/A\in[1/\sqrt{n},1/n]$ where $A$ is the area of region $\mathcal{C}$ and $n\propto T$ is the factor increase in counts, which scales linearly with exposure time $T$. A fallacy here is that the likelihood function must be zero exterior of $C^{\dagger}(X=1)$, which is a region within prior support, lest not all credible regions shrink by this factor---the latter is realistic, however, and can be viewed as consistent with numerical operation in finite-sample context.

In this work we simulated increased exposure time by assuming a scaling of $A^{\dagger}/A\approx1/2$. Numerically, given a set $\{(\bm{s},w)_{i}\}_{i=1\ldots N}$ of importance samples with weights $\{w_{i}\}$ from the density $p(M,R\;|\;\bm{d})$, we: (i) define a fiducial vector $\bm{l}$ as that of the sample reporting the highest nuisance-marginalized likelihood value; (ii) calculate $\forall i$, $\bm{s}^{\dagger}_{i}=(\bm{s}_{i}-\bm{l})/\sqrt{2}+\bm{l}$; and (iii) define $\{(\bm{s}^{\dagger},w)_{i}\}_{i=1\ldots N}$ as a set of importance samples from density $p^{\dagger}(M,R\;|\;\bm{d})$.

\end{document}